\documentclass[graybox, envcountchap, oribibl]{svmult}

\usepackage{pdflscape}

\usepackage{mathptmx}        
\usepackage{amsmath}
\usepackage{amssymb}
\usepackage{multirow}
\usepackage{color}
\usepackage{helvet}          
\usepackage{courier}         
\usepackage{dirtree}

\usepackage{makeidx}        
\usepackage{graphicx}        
\usepackage{subfig}

\usepackage{multicol}        
\usepackage[bottom]{footmisc}

\usepackage{hyperref}        
\hypersetup{colorlinks=true,urlcolor=blue}

\usepackage[misc]{ifsym}

\makeindex             

\newcommand{\notemth}[1]{\text{\scshape\tiny{#1}}}
\newcommand{\mbf}[1]{\mathbf{#1}}
\newcommand{\ee}{\mathrm{e}}
\newcommand{\ii}{\mathrm{i}}
\newcommand{\dd}{\mathrm{d}}

\newcommand{\sst}{\sin^2\!\th}
\newcommand{\cct}{\cos^2\!\th}

\newcommand{\AAEE}{\text{\AE}}

\newcommand{\al}{\alpha}

\newcommand{\ga}{\gamma}

\newcommand{\De}{\Delta}

\renewcommand\th{\theta}

\newcommand{\la}{\lambda}

\newcommand{\Sg}{\Sigma}
\newcommand{\cf}{\varphi}

\newcommand{\om}{\omega}
\newcommand{\Om}{\Omega}

\newcommand{\dl}{\partial}

\usepackage{journals}

\begin{document}

\title{Testing General Relativity with Black Hole Quasi-Normal Modes}

\author{Nicola Franchini and Sebastian H. V\"olkel}

\institute{
Nicola Franchini \at Université Paris Cit\'e, CNRS, Astroparticule et Cosmologie,  F-75013 Paris, France and CNRS-UCB International Research Laboratory, Centre Pierre Binétruy, IRL2007, CPB-IN2P3, Berkeley, US, \email{franchini@apc.in2p3.fr}
\and Sebastian H. V\"olkel (\Letter) \at Max Planck Institute for Gravitational Physics (Albert Einstein Institute),\newline D-14476 Potsdam, Germany,\newline \email{sebastian.voelkel@aei.mpg.de}
}

\maketitle

\abstract{The purpose of this chapter is to provide an overview of the exciting field of black hole quasi-normal modes and its capabilities to test general relativity in the 21st century. 
After motivating this line of research, we provide a qualitative introduction to the concept of quasi-normal modes and outline black hole perturbation theory. 
With the perturbation equations at hand, we discuss common methods to compute the quasi-normal mode spectrum and compare the advantages and disadvantages of each approach. 
We also provide an overview of possible deviations from general relativity and how they modify the quasi-normal mode spectrum of black holes from a theoretical point of view. 
We then review the rapidly evolving status of currently operating gravitational wave observatories and experimental results. 
The chapter concludes with a discussion of open problems and promising outlooks to theoretical and experimental developments. 
Central pieces that make this chapter particularly interesting, also for advanced readers, are comprehensive tables providing a detailed overview of the status of techniques to compute quasi-normal modes and methods to describe quasi-normal modes of rotating black holes beyond general relativity. 
}

\section{Introduction and Motivation}\label{sec1}

Albert Einstein's general theory of relativity~\cite{https://doi.org/10.1002/andp.19163540702} has proven to be one of modern physics' most successful theories. 
For more than 100 years, it has been outstanding in its deep concept to describe the complex interplay of time, space and matter. 
The theory has been well tested in terrestrial and solar system tests of various kind and plays an important role since the second half of the 20th century for astrophysical observations of compact objects, as well as for describing the large scale structure of the universe~\cite{Will:2014kxa}. 
In recent years, two large scale experimental efforts have opened the possibility to explore the strong and dynamical regime of gravity around black holes in unprecedented ways. 
The first one resulted in the enormous success to directly measure gravitational waves using laser interferometers, which was first achieved by the two LIGO detectors in the US~\cite{LIGOScientific:2016aoc}. 
Quickly afterwards the Virgo observatory in Italy joined and the first measurement of a binary neutron star merger was possible \cite{LIGOScientific:2017vwq}. 
Most recently, the KAGRA observatory in Japan joined the observational efforts. 
So far, almost 100 compact binary mergers have been observed~\cite{LIGOScientific:2020tif,LIGOScientific:2021sio}. 
All of them are in agreement with the coalescence of stellar mass black holes and neutron stars as described by general relativity. 
The second experimental success is the achievement to use very long baseline interferometry to construct an ``Earth-sized'' radio telescope capable of resolving horizon scale structure around the super-massive black holes M87$^*$ and Sgr A$^*$, as done by the Event Horizon Telescope (EHT) Collaboration~\cite{EventHorizonTelescope:2019dse,EventHorizonTelescope:2022wkp}. 
Complementary activities that have verified general relativity recently are for instance the high precision tracking of the S2 star around the supermassive black hole Sgr A$^*$ by the GRAVITY Collaboration~\cite{GRAVITY:2018ofz,GRAVITY:2020gka}, increasingly precise radio measurements of binary pulsar systems~\cite{Wex:2020ald}, and x-ray spectroscopy of compact binaries~\cite{Bambi:2015kza}, just to name a few of them. 
Other chapters of this book cover additional techniques and their state of the art, which is why we do not cover them in the following. 
The fact that two recent Nobel Prizes in Physics (2017 and 2020) have been awarded to black hole related topics undermine the overall significance of gravity in contemporary research. 

Besides the apparent success of general relativity, there are well-known problems that may or may not be related to the theory itself. 
Be it the unclear origin of dark matter and dark energy, or finding a generally accepted quantum theory of gravity, it might well be that more precise experiments could find deviations from general relativity. 
Although some alternative theories of gravity can be tightly constrained, many others are too complicated to work with. 
For instance, the lack of numerical relativity simulations or perturbation theory for rotating black holes in such alternative theories prevents one from having clear predictions of the strong dynamical regime of gravity. 
Without a comparable understanding and quantitative comparison of such alternatives, {\it e.g.}, in terms of gravitational wave templates, general relativity also prevails as simplest theory via Occam's razor. 

One promising avenue that puts general relativity at increasingly more challenging tests is therefore the gravitational wave emission of binary black hole mergers. 
Such events probe extreme aspects of gravity and continuous improvements of existing detectors and future detectors promise high accuracy observations that will need to be explained in detail. 
This also challenges existing approaches to treat general relativity, since more precise data also requires more precise modeling. 

One central prediction of general relativity is that all astrophysical black holes (neglecting electrical charge) are uniquely determined by their mass $M$ and angular momentum $J = a M$ via the Kerr metric \cite{Kerr:1963ud,Carter:1971zc,Robinson:1975bv}. 
In Boyer-Lindquist coordinates $x^\mu = \left(t,r,\th,\phi\right)$ it is given by
\begin{equation}\label{eq:kerr_metric}
\begin{split}
\dd s^2 = & \, - \left(1-\frac{2Mr}{\Sg}\right)\dd t^2 - \frac{4aMr\sst}{\Sg}\dd t \dd\phi + \frac{\Sg}{\De}\dd r^2 \\
& + \Sg\dd\th^2 + \left(r^2 + a^2 + \frac{2Ma^2r\sst}{\Sg} \right)\sst \dd\phi^2
\end{split}
\end{equation}
with 
\begin{align}
\De = r^2 - 2 M r + a^2, \quad \Sg = r^2 + a^2 \cct.
\end{align}
For vanishing spin $a=0$ it reduces to the Schwarzschild metric \cite{1916SPAW.......189S}
\begin{align}\label{eq:schwarzschild_metric}
\text{d}s^2 = -\left(1-\frac{2 M}{r} \right) \text{d}t^2 +  \left(1-\frac{2 M}{r} \right)^{-1} \text{d}r^2 + r^2 \text{d}\theta^2 +  r^2 \sst \text{d}\phi^2.
\end{align}
This is commonly known as the no-hair theorem and the actual validity of its hypotheses can be tested. 
As we review in this chapter, gravitational wave emission of black holes is in many regards similar to performing spectroscopy of atoms or seismology on Earth. 
The gravitational waves emitted by a perturbed black hole that settles to equilibrium are characterized by the so-called quasi-normal mode (QNM) spectrum. 
This set of frequencies and damping times, when extracted from observations, can be used to perform ``black hole spectroscopy''~\cite{Dreyer:2003bv}. 
The spectrum acts as a fingerprint of the system and via the no-hair theorem only depends on the mass and spin as the two unknown parameters. 
Any observation that allows to extract more than one complex frequency can thus in principle be used to test the hypotheses of the no-hair theorem and therefore our understanding of general relativity itself. 
Current observations are on the edge of performing such analysis. 
While there seems to be clear evidence for measurements of the $\ell=m=2$ fundamental mode ($n=0$), possible observations of overtones $n>0$ are under debate, while other works claim the detection of the $\ell=m=3$ mode. 
With more precise measurements in the upcoming observing run in 2023, one can expect exciting results in black hole spectroscopy.

The rest of this chapter is structured as follows. 
section~\ref{sec2} provides a first introduction to QNMs and the propagation of a scalar test field around the Schwarzschild black hole. 
The full problem of gravitational perturbations around Schwarzschild and Kerr black holes in general relativity is outlined in section~\ref{sec3}. 
The necessary tools to actually compute QNMs are then outlined in section~\ref{sec4}. 
Possible deviations from general relativity are discussed from theory dependent, as well as phenomenological perspectives in section~\ref{sec5}.
The confrontation between theoretical understanding and actual gravitational wave observations is presented in section~\ref{sec6}. 
This chapter concludes with a future outlook and open problems in section~\ref{sec7}. 
The key aspects are being summarized at the end of each section. 
We adopt units in which $G=c=1$.

\section{Theoretical Minimum of Quasi-Normal Modes}\label{sec2}

This section serves as a basic review of the concept of normal modes versus quasi-normal modes and the simplest black hole case. 
After reviewing basic examples from classical mechanics in section~\ref{sec2_nm_vs_qnm}, we outline the canonical example of a scalar test field in the Schwarzschild space-time in reference~\ref{sec2_boxphi}. 

\subsection{Normal Modes versus Quasi-Normal Modes}\label{sec2_nm_vs_qnm}

Although oscillation phenomena accompany us every day, it might be surprising that very similar concepts also apply to the dynamical properties of black holes. 
It is thus worth to review a few general observations, which we want to do in a very intuitive, non-technical way in the following. 
This also allows us to introduce some basic terminology that will be used throughout the rest of this review. 

We first consider the simple mechanical example of an idealized string with length $L$ of constant density $\rho$ and tension $T$. 
The dynamical properties of the string are governed by the \textit{wave equation}
\begin{align}\label{sec2_string_wave_eq}
\frac{\mathrm{d}^2\phi(t,x)}{\mathrm{d}x^2} - \frac{1}{v^2} \frac{\mathrm{d}^2\phi(t,x)}{\mathrm{d}t^2} = 0,
\end{align}
with $v^2 = T/\rho$. 
This equation can be solved by separating the temporal and spatial part via $\phi(t,x) = f(t) g(x)$, choosing an harmonic ansatz for the temporal part $f(t) \propto \exp\left(-i \omega  t \right)$ and providing \textit{boundary conditions}. 
In the following, we assume they are of Dirichlet type and correspond to fixed ends by setting $\phi(t,0) = \phi(t,L) = 0$. 
This yields the familiar result 
\begin{align}
\phi(t, x) = \sum_{n=0}^{\infty} A_{n} \exp\left[-\mathrm{i} ( \omega_n  t- k x)  \right] +  B_{n} \exp\left[-\mathrm{i} (\omega_n t+ k x)  \right],
\end{align}
where $A_n = -B_n$ are the mode amplitudes, $k = \omega_n/v$ and $\omega_n$ the \textit{normal mode} eigenvalue spectrum 
\begin{align}
\omega_n  = \frac{\pi n} {L} \sqrt{\frac{T}{\rho}},\qquad n \in \mathbb{N}.
\end{align}

Let us make some fundamental remarks in hindsight of the full problem. 
First, once the string is excited, {\it i.e.},  described by non-zero amplitudes at $t=0$ (\textit{initial data}), the oscillation goes on forever. 
This is because there is no dissipation and the mode spectrum $\omega_n$ is real. 
Second, the eigenvalue spectrum $\omega_n$ does not depend on how the system has been initially excited, but is determined by the wave equation equation~\eqref{sec2_string_wave_eq} and imposing specific boundary conditions. 
One may thus call the mode spectrum an ``\textit{intrinsic}'' property of the system, while the amplitudes encapsulate the details of the chosen initial data. 
This however does not mean, that the amplitude of each mode is independent of the properties of the system, they can in more detail be understood as a combination of intrinsic and extrinsic properties. 

Moving on to realistic oscillations, we know that they are always subject to some sort of dissipation. 
Initial excitations of a realistic string ``\textit{ringdown}'' to equilibrium by loosing energy, {\it e.g.}, by couplings to the surroundings via sound waves and internal dissipation. 
This damping phenomena controlled by the presence of a non-zero imaginary part gives rise to the term \textit{quasi-normal modes}. 

To connect mechanical system oscillations with those of black holes, let us briefly mention neutron stars next. 
The complexity of neutron star oscillations is overwhelming and ultimately connects nuclear physics with strong field gravity. 
For some classical reviews on the topic we refer to references~\cite{Kokkotas:1999bd,Nollert:1999ji}. 
As for normal stars, there are different types of oscillation modes that are characterized by the restoring forces. 
For an idealized star in Newtonian gravity, in the absence of any internal dissipation mechanisms, oscillations can indeed be described by normal modes. 
In general relativity, there is a fundamental difference due to the metric. 
It is known that non-radial oscillations couple to the metric, thereby emit gravitational waves, and are thus one novel mechanism to remove energy. 

Although general relativity and neutron star physics yield very complicated equations, one can understand some of the basic properties of emitted waves with simplified models. 
One such model has been proposed by Kokkotas and Schutz in reference~\cite{Kokkotas:1986gd} by coupling two different types of strings. 
Here, the basic idea is that the oscillations of the neutron star can be captured by those of a finite string, which is then coupled to a semi-infinite string, corresponding to the space-time and its gravitational waves. 
An analytic treatment of the problem reveals different types of eigenvalues, which depend on how strong the two strings are coupled. 
There are some eigenvalues that are very similar to those of the uncoupled finite string, but now include an exponentially small imaginary part. 
Depending on the location where the strings are coupled, there are also modes at nodal points that are not coupled at all. 
Some modes are strongly damped, which cannot be identified with the modes of the finite string. 
The latter type of modes is qualitatively more related to the QNMs of black holes, which we discuss for the simplest case in the next section. 
Here the analogue system would be more accurately described by a single infinite string that cannot support any sort of long lived modes.

\subsection{Scalar Field around a Schwarzschild Black Hole}\label{sec2_boxphi}

As first introduction to black hole QNMs, one often finds the study of the propagation of a scalar test field on a fixed Schwarzschild background~\cite{Maggiore:2018sht}. 
This simple problem already highlights many of the key features and challenges when studying gravitational perturbations. 
The action of a massless scalar field $\varphi$ in the background space-time $g_{\mu \nu}$ is given by
\begin{align}\label{eq:scalar_action}
S = \int \text{d} x^4 \sqrt{-g} \frac{1}{2} \left( \partial_\mu  \varphi  \right) \left( \partial^\mu \varphi \right) \,.
\end{align}
From the variation of this action, one finds the Klein-Gordon equation of motion, in the form of the wave operator on a curved background
\begin{align}
\Box \varphi 
= \left(- g \right)^{-1/2} \partial_\mu \left[\left(- g \right)^{1/2} g^{\mu \nu} \partial_\nu \right] \phi 
=0.
\end{align}

For a general metric the previous equations are non-trivial to study. 
However, since the Schwarzschild metric is static and spherically symmetric, the equations can be simplified significantly. 
Writing out the sum over all indices one finds that the radial-time and angular parts of the equations can be separated via
\begin{align}
\varphi = \sum_{\ell,m} \frac{u(t,r)}{r} Y_{\ell m}(\theta, \phi).
\end{align}
A closer look at the angular part shows that it is solved by the spherical harmonics, defined from the associated Legendre polynomials $P_{\ell m}(x)$ as
\begin{align}\label{eq:SphericalHarmonics}
    Y_{\ell m}(\th,\phi) = \sqrt{\frac{(2\ell -1)}{4\pi}\frac{(\ell - m)!}{(\ell+m)!}} P_{\ell m}(\cos\th)\ee^{\ii m \phi},
\end{align}
and the separation constant is known to be $\ell(\ell+1)$. 
The explicit time dependency can be replaced by Fourier transforming the equations
\begin{align}\label{def_fourier}
u(t,r) = \int_{-\infty}^{+\infty} \frac{\dd\omega}{2\pi} \tilde{u}(\omega, r) \ee^{-\ii \omega t}.
\end{align} 
Finally, the separated time-radial part can be further simplified by introducing the so-called tortoise coordinate
\begin{align}\label{tortoise}
r^*(r) = r + 2\,M \log\left(\frac{r}{2\,M}-1\right).
\end{align} 
Since the angular part is already solved, the final problem that needs to be solved is of the same form as the time independent Schr\"ordinger equation
\begin{align}\label{sec2_boxphi_wave}
\frac{\text{d}^2}{\text{d} {r^*}^2} \Psi  + \left[\omega^2 - V_\ell(r) \right]\Psi = 0,
\end{align}
with $\Psi \equiv \tilde{u}$. 
The effective potential $V_\ell(r)$ is given by
\begin{align}\label{sec2_boxphi_V}
V_\ell(r) = \left(1- \frac{2M}{r} \right) \left( \frac{\ell(\ell+1)}{r^2} + \frac{2M}{r^3} \right).
\end{align}
QNMs $\omega_n$ are defined as purely outgoing waves at $r \rightarrow \infty$ (no incoming radiation from infinity) and purely ingoing waves at the horizon $r \rightarrow 2 M$ (no outgoing radiation from the black hole). 
Imposing these boundary conditions defines the QNM spectrum, but it is not possible to find it in terms of simple analytic functions. 
Although equation~\eqref{sec2_boxphi_wave} with the potential equation~\eqref{sec2_boxphi_V} might not appear very complicated for a numerical study, incorporating the boundary conditions and computing the QNM spectrum is not trivial. 
The reason is that there are two asymptotic solutions for the radial equation, one that is exponentially decaying and one that is exponentially growing. 
The QNM boundary conditions correspond to the exponentially growing modes, which makes their actual treatment in numerical studies challenging. 
Without going into more details here, it is intuitively clear that using a standard approach, {\it e.g.}, the shooting method, which integrates the perturbation for a given guess of $\omega$ inwards from far away is numerically unstable. 
It turns out that the QNM spectrum of black holes is very different from commonly known ones of mechanical systems. 
Black hole QNM overtones first decrease in their frequency, before they increase again and asymptote to a constant value, while their imaginary part grows at a constant rate. 
For a quantitative description we refer the interested to Fig.~5 in reference~\cite{Berti:2009kk}. 
An insightful interpretation of the Schwarzschild QNM spectrum, which can be further related to possible quantum mechanical properties of black holes, has been discussed by Maggiore in reference~\cite{Maggiore:2007nq}. 
We review methods to compute the QNM spectrum in section~\ref{sec3_methods_qnms}. 

A final remark on how the picture changes beyond spherical symmetry. 
For the Kerr metric, as we will see in section~\ref{sec:Teukolsky}, it is still possible to decouple the equations by using spheroidal harmonics, but it is in general not possible to decouple the equations for any axial symmetric background. 
The coupled equations then have either to be studied numerically or with perturbative techniques if the coupling is weak, see references~\cite{Cano:2020cao,Ghosh:2023etd}. 

\begin{svgraybox}
In this section we discussed:
\begin{itemize}
\item \textit{normal modes} as real eigenvalues that are a property of idealized systems that do not loose energy, {\it e.g.},  an idealized string;
\item \textit{quasi-normal modes} as complex eigenvalues that appear with dissipation, {\it e.g.}, a finite string coupled to a semi-infinite string;
\item that a scalar test field in the Schwarzschild space-time gives rise to a qualitatively similar problem, and its eigenvalue spectrum is described by QNMs.
\end{itemize}
\end{svgraybox}

\section{Quasi-Normal Modes in General Relativity}\label{sec3}

In this section we review the derivation of gravitational QNMs in general relativity. 
We start in section~\ref{sec3_sch} with metric perturbations in spherical symmetry and then discuss the axial symmetric case in section~\ref{sec:Teukolsky}.

\subsection{Metric Perturbations of Schwarzschild Black Holes}\label{sec3_sch}

In the following we review the key steps to derive the equations governing the perturbations of the metric for the Schwarzschild black hole. 
They starting point is to consider an ansatz for the metric $g_{\mu \nu}$
\begin{align}
g_{\mu \nu} = \bar{g}_{\mu \nu} + h_{\mu \nu},
\end{align}
where $ \bar{g}_{\mu \nu}$ is the Schwarzschild metric given in~\ref{eq:schwarzschild_metric} and $h_{\mu \nu}$ a small perturbation. 
In contrast to the previous case of a scalar field on a fixed background, it is now necessary to study the full Einstein field equations in vacuum, which in the reduced form are simply given by
\begin{equation}\label{eq:reducedEinstein}
    R_{\mu \nu} = 0 .
\end{equation}
Inserting the ansatz in the above equations and keeping only terms up to linear order yields the standard form of the gravitational perturbation equations. The most general perturbation $h_{\mu \nu}$ is a matrix of 10 independent functions that depend on all the coordinates of the spacetime. A way to understand how these functions look like is to see how they transform under rotations. Without loss of generality, we notice that $h_{00}$ must transform under rotations as a scalar, $h_{0i}$ as a 3-dimensional vector and $h_{ij}$ as a rank-2 tensor. According to this prescription, the decomposition can be performed as follows~\cite{Bardeen:1980kt}
\begin{align}
    h_{00} & = \phi_0 ,\\
    h_{0i} & = \nabla_i \phi_1 + v_i ,\\
    h_{ij} & = \phi_2 \bar{g}_{ij} + \left(\nabla_i \nabla_j  -\frac{1}{3}\bar{g}_{ij} \nabla^2 \right) \phi_3 + \nabla_{(i} w_{j)} + S_{ij} ,
\end{align}
where $\phi_0$, $\phi_1$, $\phi_2$, $\phi_3$ are scalars under rotation (one degree of freedom each), $v_i$ and $w_i$ are divergence-free vectors (two degrees of freedom each) and $S_{ij}$ is a transverse, traceless, symmetric rank-2 tensor (two degrees of freedom). 
The definition of $h_{ij}$ is such that is manifest a splitting among the trace-free part (from second to fourth term) and the rest. The total number of degrees of freedom is ten, as expected.

\begin{figure}
\centering
\includegraphics[width=\linewidth]{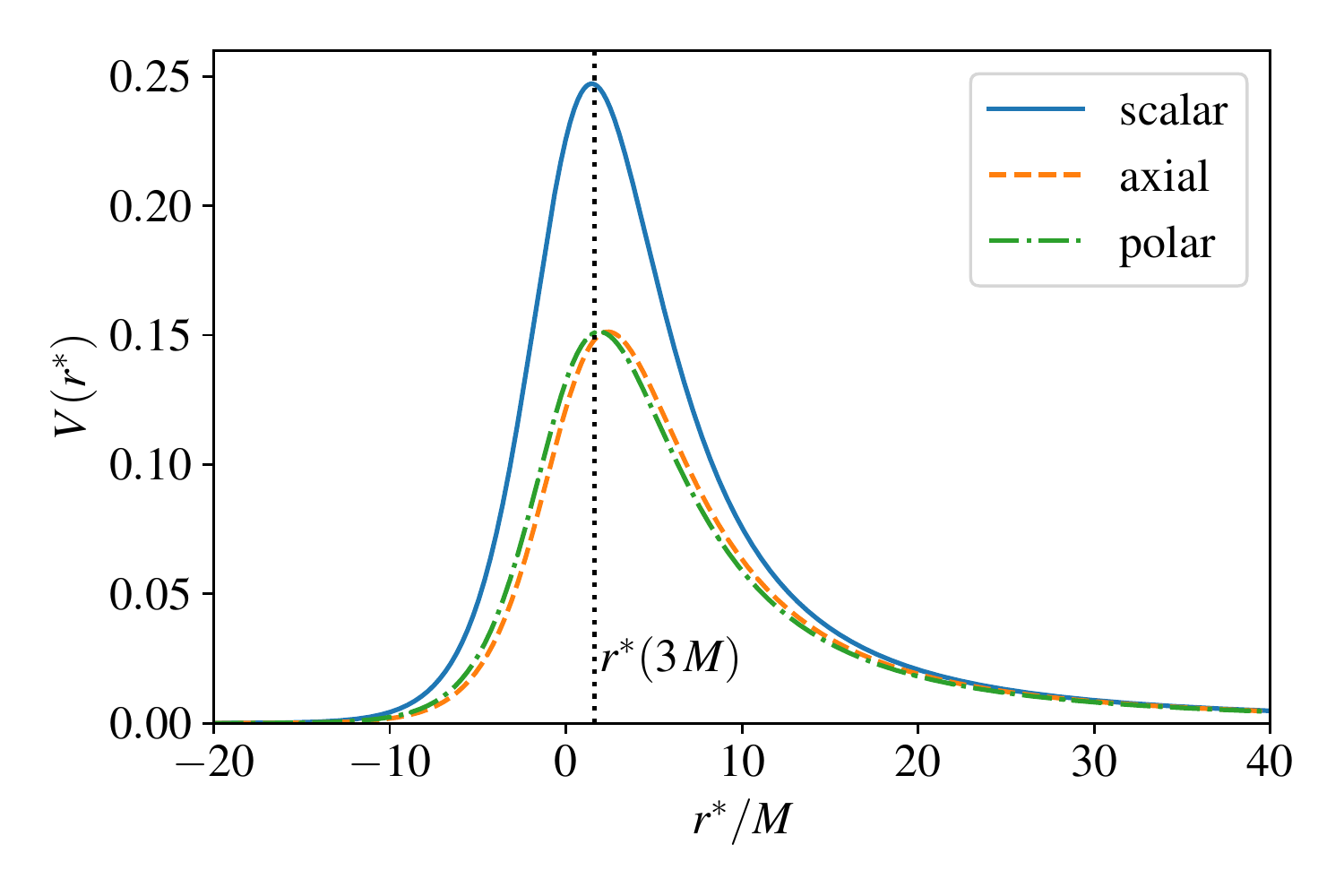}
\caption{Scalar, axial and polar potentials for $\ell=2$. The black dotted line indicates the location of the light-ring at $r_*=3M$, which coincides approximately with the potential maxima.}
\label{fig:potentials}
\end{figure}

From now on, we restrict to Schwarzschild coordinates $(t,r,\th,\phi)$. 
A convenient basis to express properties of rank-0, 1 and 2 tensors under rotations is represented by scalar, vector and tensor spherical harmonics. 
From the scalar spherical harmonics defined in equation~\eqref{eq:SphericalHarmonics}, one can define the vector and tensor harmonics as follows
\begin{align}
    \text{vector:} \quad & P^a_{\ell m} = \left( \dl_\th Y_{\ell m}, \dl_\phi Y_{\ell m} \right), \quad  A^a_{\ell m} = \left( -\frac{\dl_\phi Y_{\ell m}}{\sin\th} , \sin\th \, \dl_\th Y_{\ell m} \right), \\
    \text{tensor:} \quad & P_{\ell m}^{ab} = 
    \begin{pmatrix}
    W_{\ell m} & X_{\ell m } \\
    * & -\sst \, W_{\ell m}
    \end{pmatrix},  \quad
    A_{\ell m}^{ab} = 
    \begin{pmatrix}
    -\frac{X_{\ell m}}{\sin\th} & \sin\th\,W_{\ell m } \\
    * & \sin\th\, X_{\ell m}
    \end{pmatrix} ,
\end{align}
where asterisks denote symmetric matrix element and
\begin{align}
    X_{\ell m} & = \left(2\dl_\th\dl_\phi - 2\cot\th \dl_\phi\right) Y_{\ell m}, \\    W_{\ell m} & = \left( \dl_\th^2 - \cot\th \dl_\th - \frac{1}{\sst}\dl_\phi^2 \right) Y_{\ell m}.
\end{align}
The tensors denoted with $P$ gain a factor $(-1)^\ell$ under parity inversion $\th\rightarrow\pi-\th$ and $\phi\rightarrow\phi + \pi$ hence are said to be polar or even. On the other hand, the tensors denoted by $A$ get a factor $(-1)^{\ell+1}$ under parity transformation and are called axial or odd. 
With these definitions of spherical harmonics one can construct ten independent basis vectors $\mbf{t}^a_{\ell m,\mu\nu}$ that respect the scalar-vector-tensor decomposition outlined before. 
Their explicit form can be found in~\cite{Maggiore:2018sht}. 
Since we are considering the Schwarzschild metric as our background and it is invariant under parity inversion, it means that the perturbations will naturally split into axial and polar modes that do not couple to each other. 
Following~\cite{Maggiore:2018sht}, the even modes under parity are those labelled by the index $a = tt,\,Rt,\,L0,\,T0,\,Et,\,E1,\,E2$ while the odd modes are $a = Bt,\,B1,\,B2$. 
The metric perturbation $h_{\mu\nu}$ can be now expanded in this basis as follows
\begin{equation}
    h_{\mu\nu} = \sum_{a} \sum_{\ell m } h^a_{\ell m}(t,r)\mbf{t}^a_{\ell m,\mu\nu}(\th,\phi)\,
\end{equation}
where the sum over $a$ is intended over the 10 indices. This naturally splits the perturbation metric into
\begin{equation}
    h_{\mu\nu} = h_{\mu\nu}^\notemth{polar} + h_{\mu\nu}^\notemth{axial}.
\end{equation}
So far, we did not make any gauge assumption. 
Any infinitesimal gauge transformation can be used to get rid of up to four degrees of freedom in the metric. 
The most common gauge used in black hole perturbation theory is the Regge-Wheeler (RW) gauge, which sets to zero the modes corresponding to the indices $a = Et,\, E1,\, E2,\, B2$. 
More details can be found in~\cite{Maggiore:2018sht}. 
In this gauge, we have the following representation for the odd and even metric perturbations
\begin{align}
    h_{\mu\nu}^\notemth{odd} & = \sum_{\ell m}
    \begin{pmatrix}
    0 & 0 & h_0^{\ell m}(t,r)/\sin\th \dl_\phi & h_0^{\ell m}(t,r) \sin\th \dl_\th \\
    0 & 0 & h_1^{\ell m}(t,r)/\sin\th \dl_\phi & h_1^{\ell m}(t,r) \sin\th \dl_\th \\
    * & * & 0 & 0 \\
    * & * & 0 & 0 \\
    \end{pmatrix}Y_{\ell m} ,\\
    h_{\mu\nu}^\notemth{even} & = \sum_{\ell m}
    \begin{pmatrix}
    \bar{g}_{tt} H_0^{\ell m}(t,r) & H_1^{\ell m}(t,r)        & 0 & 0 \\
    *                   & \bar{g}_{rr} H_2^{\ell m}(t,r) & 0 & 0 \\
    0                   & 0                   & r^2 K^{\ell m}(t,r) & 0 \\
    0                   & 0                   & 0 & r^2 \sst K^{\ell m}(t,r) \\
    \end{pmatrix}Y_{\ell m} .
\end{align}
Finally, we perform a Fourier decomposition of all the functions
\begin{equation}
    h^a(t,r) = \int_{-\infty}^{+\infty} \frac{\dd\om}{2\pi}\tilde{h}^a(\om,r)\ee^{-\ii \om t}
\end{equation}
Given this {\it ansatz}, one can compute the Ricci tensor $R_{\mu \nu}$, plug it into equation~\eqref{eq:reducedEinstein} and extract the first-order contribution. 
The calculation is lengthy, but it can be carried out easily with computer algebra systems like \textsc{Maple} or \textsc{Mathematica}. 
The ten equations obtained in this way naturally split into three odd-parity equations and even-parity equations. 
They are not all independent, and in fact, after some manipulation, they can be recast into two equations, notably known as RW equation for the odd sector, here denoted with $(-)$ label, and Zerilli equation for the even sector, here denoted with $(+)$ label
\begin{align}\label{eq:RW_Zerilli}
    \frac{\dd^2 \Psi^{(\pm)}_\ell}{\dd r_*^2} & + \left[ \om^2 - V^{(\pm)}(r) \right] \Psi^{(\pm)}_\ell = 0 
\end{align}
where
\begin{align}
    V^{(-)}(r) & = \left(1 - \frac{2M}{r} \right)\left( \frac{\ell(\ell+1)}{r^2} - \frac{6M}{r^3} \right) \label{eq:RW}\\
    V^{(+)}(r) & = \left(1 - \frac{2M}{r} \right)\left( \frac{\ell(\ell+1)}{r^2} - \frac{6M}{r^3}\frac{r^2\la(\la+4) + 12M r - 12M^2}{(6M+r\la)^2} \right) \label{eq:Zerilli}
\end{align}
and $\la = \ell(\ell+1) -2$. 
Note that $V^{(-)}(r)$ is very similar to the scalar field potential shown in \eqref{sec2_boxphi_V} although the metric perturbations are much more involved. In Fig.~\ref{fig:potentials} we show the $\ell=2$ case for the scalar, RW and Zerilli potentials as function of the tortoise coordinate $r^*(r)$ defined in~\eqref{tortoise}. 
All three potentials are qualitatively very similar, {\it e.g.}, they have a single maximum around $r_*\approx 3M$ and decay monotonously to zero towards the horizon (exponentially) and towards radial infinity ($\sim 1/r^2$). 
For large values of $\ell$, the height of all the potential barrier increases, but the location of each maximum gets closer to the location of the lightring at $r_*=3M$. 

We wrote equations~\eqref{eq:RW}--\eqref{eq:Zerilli} such that it appears evident that in the limit $\ell\rightarrow\infty$ the two potentials are identical, and dominated by the term $\ell(\ell+1)/r^2$. 
It is evident, that in this limit, also known as {\it eikonal} or {\it geometric-optic} limit, the QNM potential is identical to the effective potential of a massless particle in geodesic motion around a Schwarzschild. 
This can be used to approximate the QNMs and is discussed in more detail in section~\ref{methods_eikonal}. 

A different way to obtain the RW and the Zerilli equations was pointed out by Moncrief, who implemented a gauge invariant derivation starting from the Arnowitt-Deser-Misner decomposition of the spacetime~\cite{Moncrief:1974am}. Another striking aspect that arises in the computation of QNMs is the {\it isospectrality} of the RW and of the Zerilli equations. 
This feature refers to the fact that the spectrum of the two potentials is identical, albeit they are explicitly different. 
The solution to this problem was found by Chandrasekhar~\cite{Chandrasekhar:1985kt}, who realised that the RW and the Zerilli equations are linked by a Darboux transformation~\cite{Glampedakis:2017rar}. Even from Fig.~\ref{fig:potentials}, despite isospectrality between the axial and polar potentials, one may not be surprised that they are actually very similar, almost as if they are just ``shifted'' by a small offset.

\subsection{Teukolsky's Approach for Kerr Black Holes}\label{sec:Teukolsky}
Obtaining an equation for gravitational perturbations of the Kerr metric~\eqref{eq:kerr_metric} involves a more convoluted computation that makes use of the Newman-Penrose decomposition in null tetrads of the metric. 
Without entering in the details of the calculation, which are extensively covered in the literature~\cite{Teukolsky:1973ha,Chandrasekhar:1985kt,Maggiore:2018sht}, we summarize the main result, which is the Teukolsky master equation
\begin{equation}\label{eq:teukolsky}
\begin{split}
    & \left[\frac{\left(r^2 + a^2\right)^2}{\De} - a^2 \sst \right] \dl^2_t \psi^{(s)} 
    + \frac{4aMr}{\De} \dl_t \dl_{\phi}\psi^{(s)} 
    + \left[\frac{a^2}{\De} - \frac{1}{\sst} \right] \dl^2_{\phi} \psi^{(s)} \\
    & - \De^{-s} \partial_r \left( \De^{s+1} \partial_r \psi^{(s)} \right) - \frac{1}{\sin{\th}} \dl_{\th} \left( \sin{\th} \dl_{\th} \psi^{(s)} \right) \\
    & - 2s \left[\frac{a \left(r-M \right)}{\De}+\ii \frac{\cos{\th}}{\sst} \right] \dl_{\phi} \psi^{(s)}
    + (s^2 \cot{\th}^2 -s ) \psi^{(s)} \\
    & -2s \left[ \frac{2M(r^2-a^2)}{2\De} - r - \ii a \cos{\th} \right] \dl_t \psi^{(s)} 
     = 0 .
\end{split}
\end{equation}
The Teukolsky equation is valid for perturbations of different spin-weight, namely, scalars ($s=0$), spinors ($s=\pm1/2$), vectors ($s=\pm1$) and, the case of interest for gravitational waves, tensor perturbations ($s=\pm2$). It turns out that also for the Teukolsky equation one can decouple the radial and the angular part, provided that a suitable basis for the angular wavefunctions is chosen. Thus, we consider the following decomposition
\begin{equation}
    \psi^{(s)} = \int_{-\infty}^{+\infty} \frac{\dd \om}{2\pi} \sum_{\ell m} R^{(s)}_{\ell m}(\om,r) S^{(s)}_{\ell m}(\om,\th) \ee^{-\ii(\om t - m \cf) } 
\end{equation}
where the Fourier decomposition was taken into account as well.
For the angular part one gets
\begin{align}\label{eq:SpheroidalHarmonics}
    \frac{1}{\sin\th} \dl_\th\left( \sin\th \dl_\th S_{\ell m}^{(s)} \right) + & \bigg[ a^2 \om^2 \cct - \frac{m^2}{\sst} - 2 s a \om \cos\th \notag \\
    & - 2 s m \frac{\cos\th}{\sst} - (s^2 \cot^2\!\th - s) + \la_{\ell m}^{(s)} (\om) \bigg] S_{\ell m}^{(s)} = 0 ,
\end{align}
where $\la_{\ell m}^{(s)}$ is the separation constant and it depends on the frequency $\om$. The functions $S_{\ell m}^{(s)}$ are known as {\it spin-weighted spheroidal harmonics}, and in the zero spin limit they reduce to the $\th$ component of the spin-weighted spherical harmonics $Y_{\ell m}(\th,\cf) \ee^{- \ii m \cf}$. The functions $S_{\ell m}^{(s)}$ are, in general, not known analytically, but they can be computed numerically for each different $s,\ell,m,a$ and $\om$ with different methods. The radial equation must satisfy the following equation
\begin{align}\label{eq:Teukolsky_Radial}
    \De^{-s} \dl_r & \left( \De^{s+1} \dl_r \right) R_{\ell m}^{(s)} \notag \\
    + & \bigg[ \frac{K^2 - 2 \ii s(r - M)K}{\De} - a^2 \om^2 + 2 m a \om +  4\ii s \om r  - \la_{\ell m}^{(s)}\bigg] R_{\ell m}^{(s)} = 0,
\end{align}
with
\begin{equation}
    K = \left(r^2 + a^2 \right) \om - a m .
\end{equation}
In order to find the spectrum of QNMs for a Kerr black hole of spin $a$, one needs to solve simultaneously equations~\eqref{eq:SpheroidalHarmonics}--\eqref{eq:Teukolsky_Radial}, imposing appropriate boundary conditions. For the radial functions $R_{\ell m}^{(s)}$ one wants to impose that the are purely ingoing at the event horizon and purely outgoing at spatial infinity
\begin{equation}
    \lim_{r\to\pm r_*} R_{\ell m}^{(s)}(r) \propto \ee^{\pm \ii \om r_*},
\end{equation}
whereas the spin-weighted spheroidal harmonics are taken to be regular on the interval $\th \in [0,\pi]$. The spectrum of QNMs is composed of infinite modes for each $\ell$ and $m$ selected. As in the spherically symmetric case these modes are ordered in overtones $n$ from the most long-lived onwards, {\it i.e.}, smaller $n$ correspond to larger damping time $\tau_n = 1/ \textrm{Im}(\om_n)$. The difference with the spherically symmetric case is that the index $m$ is not degenerate, due to the rotation. Anyway, parity symmetry around the $\th$ plane ensures that
\begin{equation}
    \om_{n\ell m} = - \left( \om_{n \ell - m}' \right)^*.
\end{equation}
We conclude by closing the picture of the relation between the RW and Zerilli equations with the Teukolsky equation. Indeed, Chandrasekhar realised that both the RW and Zerilli can be generated from the Bardeen-Press equation (the $a=0$ limit of the Teukolsky equation)~\cite{Chandrasekhar:1985kt} with a generalization of the already encountered Darboux transformation~\cite{Glampedakis:2017rar}. 

In the next section we will see some of the most common methods for the computation of QNMs for a given perturbation equations.

\begin{svgraybox}
In this section we discussed:
\begin{itemize}
\item how the gravitational perturbations of the Schwarzschild black hole can be derived with metric perturbation theory;
\item how the gravitational perturbations of the Kerr black hole can be derived with the Newman-Penrose formalism;
\item that there are many qualitative similarities with respect to the test scalar field case.
\end{itemize}
\end{svgraybox}

\section{Computation and Other Aspects of Quasi-Normal Modes} \label{sec4}

In section~\ref{sec3_methods_qnms} we first provide a review of the most common techniques to compute the QNM spectrum from the perturbation equations of the previous sections. 
Different aspects related to QNMs beyond their spectrum are discussed afterwards. 
In section~\ref{sec3_relevance} we address the question of the significance of QNMs for perturbed black holes. 
The limitations of linear perturbation theory for a ringdown analysis are discussed in section~\ref{sec3_linear_vs_nonlinear}.

\subsection{Techniques for Quasi-Normal Mode Computations}\label{sec3_methods_qnms}

Compared to the full field equations of general relativity, the wave equations for linearized perturbations look significantly less involved. 
It may thus seem surprising that the actual computation of QNMs can be a rather delicate task. 
Depending on the specific objectives in mind, there are different types of methods available, which we summarize qualitatively in the following. 
Note that the following list is not complete, but merely serves the purpose to get an overview of the different directions. 
More information can also be found in references~\cite{Kokkotas:1999bd,Nollert:1999ji,Berti:2009kk,Konoplya:2011qq,Pani:2013pma}. 
We provide an overview of the commonly used methods in Table~\ref{table_qnm_methods} that indicates the advantages and disadvantages of each method.

\subsubsection*{Eikonal Limit}\label{methods_eikonal}

One of the most common approximations in the calculation of QNMs is the so-called geometric-optic or eikonal limit in the form that states the correspondence of QNMs to properties of geodesics. 
To avoid possible confusion with the popular higher-order WKB method, which is sometimes also referred to as eikonal approximation (see later), let us say that the two methods share some assumptions, but they require different ingredients and differ in their applicability. 
The eikonal limit is usually utilized by \textit{assuming} that in the large $\ell$ limit there is a direct correspondence between the perturbation equation potential and the potential for geodesics in the same spacetime. 
In particular, in this limit the QNMs can be related to the orbital frequency $\Omega$ of null rays at the unstable photon orbit and the Lyapunov exponent of the orbit $\gamma_L$. 
This equivalence was first noticed by Goebel~\cite{1972ApJ...172L..95G} based on the results of the large $\ell$ computation of the real part of the frequency of long-ranged perturbations performed by Press~\cite{1971ApJ...170L.105P}, showing that they are proportional to the orbital frequency $\Om$ of null rays at the unstable circular orbit. 
This result was later generalised by Ferrari and Mashhoon to link the imaginary part to the Lyapunov exponent of the orbit $\ga_L$~\cite{Ferrari:1984zz,Ferrari:1984ozr}
\begin{equation}
    \om \simeq \left(\ell + \frac{1}{2}\right)\Om - \ii \ga_L\left(n + \frac{1}{2}\right).
\end{equation}
Further generalizations of this formula appeared in the context of charged and rotating black holes in general relativity~\cite{Mashhoon:1985cya,Dolan:2010wr,Yang:2012he}, black holes in higher dimensions~\cite{Cardoso:2008bp} and black holes in theories beyond general relativity~\cite{Glampedakis:2017dvb,Glampedakis:2019dqh,Silva:2019scu,Bryant:2021xdh}. 
In the absence of knowing the full gravitational perturbation equations for a comparison, it is unclear under which assumptions the geometric-optic correspondence is a property valid for extensions of general relativity (see, {\it e.g.}, references~\cite{Khanna:2016yow,Konoplya:2017wot,Konoplya:2022gjp}, for cases where it is not valid or limited).

\subsubsection*{Inverted Potential Method}

Full analytic, but approximate results for some part of the QNM spectrum may be obtained in case the perturbation equation is in the form of the time independent Schr\"odinger equation with a simple potential term. 
The idea of the inverted potential method is to match the parameters of a simpler potential whose spectrum can be computed analytically. 
It was first introduced to black holes by Mashhoon in reference~\cite{Mashhoon:1982im} and extended in references~\cite{BLOME1984231,Ferrari:1984ozr,Ferrari:1984zz,Churilova:2021nnc}. 
It is called inverted potential method, because a complex coordinate transformation allows one to use the analytic form of the bound state spectrum of the corresponding potential well, to provide the exact QNM spectrum of the matched potential barrier. 
Potentials that are commonly used are the harmonic oscillator potential and the P\"oschl-Teller potential. 
In both cases one matches the approximate potential by demanding that its value and its second derivative agree at the maximum of the barrier with the true potential. 
For the RW potential it can be done analytically. 
Since the matching of the potential is good around the peak, but becomes less accurate further away, the method is useful to approximate the fundamental mode, but becomes less accurate for overtones. 
An explicit frequency dependency of the potential can also be studied in an approximate way, see the slow spin calculation presented in the original work \cite{Ferrari:1984ozr,Ferrari:1984zz}. 
The advantages of the method are that it is very easy to use and can provide analytic, but approximate results. 
The disadvantages are that the method requires the Schr\"odinger wave equation with a potential that is close to the analytically solvable one and cannot straight forwardly be improved to achieve higher accuracy. 
Using the WKB method it was shown that this method can be improved significantly in references~\cite{Zaslavsky:1991ug,Hatsuda:2019eoj,Matyjasek:2019eeu}, while a full numerical approach was presented in reference~\cite{Volkel:2022ewm}.

\subsubsection*{Higher-Order WKB Method}\label{method_wkb}

The Wentzel-Kramers-Brillouin (WKB) method is a well-known approach to study approximate solutions of differential equations and has been widely applied to the Schr\"odinger equation in quantum mechanics \cite{Dunham:1932zz}, but also many other applications, see reference~\cite{bender1999advanced} for a standard reference. 
In a series of papers~\cite{Schutz:1985km,Iyer:1986np,Iyer:1986nq,Kokkotas:1988fm,Seidel:1989bp,Konoplya:2003ii,Matyjasek:2017psv} it has been shown that the general WKB ansatz can be used to derive very specific analytic approximations of the QNM spectrum of potential barriers with a single maximum and suitable asymptotic behaviour 
\begin{align}\label{wkb_higherorder}
    \omega_n^2 = V^{(0)} - \ii \sqrt{-2V^{(2)}}\left(n+\frac{1}{2} \right) + \sum_i{\tilde{\Lambda}_i(n)} \,.
\end{align}
Here the numbers in the brackets indicate derivatives that are taken with respect to the tortoise coordinate and evaluated at the peak of the potential. 
Increasing the order of the WKB ansatz allows in principle to increase the accuracy of the QNM spectrum, but it must be noted that the WKB series is asymptotic. 
This means that increasing the order does not always lead to more accurate results, but it is usually very good. 
More recently, the higher-order WKB method has been extended to very high order and further improved by using Pad\'e approximants and Borel series to resum the asymptotic series for better results \cite{Hatsuda:2019eoj,Matyjasek:2019eeu}. 
In references~\cite{Zaslavsky:1991ug,Hatsuda:2019eoj,Matyjasek:2019eeu} the higher-order WKB method results are obtained by studying the bound state problem, see previous paragraph. 
The method is very popular because the analytic result is completely characterized by the Taylor expansion coefficients of the potential around its peak (expressed in the tortoise coordinate) and can be computed rather easily. 
Although the expressions become lengthy with increasing WKB order, the accuracy for $n \lesssim  l$ is usually very high, see in particular reference~\cite{Konoplya:2019hlu} for a recent overview. 
However, the standard method is limited to the Schr\"odinger equation with single maximum potential barriers with standard QNM boundary conditions. 
An extension to include a possible frequency dependency in the potential has been presented using the WKB related Bohr-Sommerfeld quantization rule \cite{Kokkotas:1991vz}, but it requires more care and is less trivial than the standard case. 
Related to this approach is also the phase integral method that has been applied to black holes in references~\cite{Froeman:1992gp,Andersson:1992scr,NAndersson_1993,Andersson:1995vi}. 
In comparison to the standard higher-order WKB method, the phase integral method is overall more accurate, and it can even be adjusted to describe strongly damped modes. 
However, due to the complex integration, it is more difficult to handle than the simple-to-calculate equation~\eqref{wkb_higherorder}. 
Finally, the WKB method can be extended to multi-field cases with coupled systems of equations~\cite{Hui:2022vov}.

\subsubsection*{WKB Methods for Exotic Compact Objects}

While some of the other methods can also be applied to exotic compact objects\footnote{The standard higher-order WKB method cannot, because it is based on different boundary conditions at the horizon. 
Still, results obtained this way can give a good description of the early ringdown modes before echoes, but those are not anymore eigenvalues of the system.}, there are a few easy-to-use, low-order WKB methods to compute their so-called quasi-stationary or trapped modes spectra. We refer to reference~\cite{2013waap.book.....K} for a textbook overview of such methods and references~\cite{Cardoso:2014sna,Volkel:2017ofl,Jayawiguna:2022ftj} for some applications. 
For ultra-compact, horizonless objects, {\it e.g.}, constant density stars, gravastars, non-zero reflection at a modified horizon scale
ultra compact stars and gravastars, there can be new type of modes and we refer to section~\ref{sec4_exotic_compact_objects} for more information on these objects. 
Although these characteristic modes are absent in asymptotically flat, Schwarzschild/Kerr black holes, similar situations also arise for AdS black holes~\cite{Festuccia:2008zx}. 
One of the typical WKB results is the generalized Bohr-Sommerfeld rule
\begin{align}\label{sec_methods_BS}
\int_{x_0}^{x_1} \sqrt{\omega_n^2 - V(x)} \text{d} x = \pi \left(n+\frac{1}{2} \right) - \frac{i}{4} \exp\left(2 i \int_{x_1}^{x_2} \sqrt{\omega_n^2 - V(x)} \text{d} x  \right),
\end{align}
where $x_0, x_1, x_2$ are the turning points of a suitable potential defined by the roots of the integrand ($\omega^2 = V(x)$). 
The method is qualitatively valid for modes with $\text{Im}(\omega_n) \ll \text{Re}(\omega_n)$ and can be expanded to yield the classical Bohr-Sommerfeld rule (without the rhs. imaginary part of eq.~\eqref{sec_methods_BS}) and the Gamow formula~\cite{Gamow:1928zz}. 
Approximate relations of the form of eq.~\eqref{sec_methods_BS} have furthermore the rather unique advantage that they can be ``inverted'' to constrain properties of the potential starting from the QNMs, see section~\ref{sec4_inverse} for a discussion about the inverse problem.

\subsubsection*{Leaver Method}

The Leaver method~\cite{Leaver:1985ax} is usually considered as the most precise method to compute QNMs with almost arbitrary precision. 
The main idea is to have a guess for the eigenfunction that captures at the same time the asymptotic behaviour of the solution at both singular points of the equation. The next step is to identify the function that vanishes at the black hole horizon and define a power series of this function. By multiplying the initial guess by this power series one gets the ansatz to be inserted in the perturbation equation. 
If one succeeds to find a recurrence relation for the coefficients of the series the solution can be written in terms of continued fractions. 
This allows one to find high accuracy QNMs up to a moderate range of overtones $n$. This limitation can be solved by a proposal of Nollert by inspecting the asymptotic behaviour of the expansion~\cite{Nollert:1993zz} which was further generalized in reference~\cite{Zhidenko:2006rs}. 
The method is not limited to the standard Schr\"odinger equation, but can also be generalized to some coupled system of equations~\cite{Rosa:2011my,Pani:2012bp,Pani:2013pma,Volkel:2022aca}. 
Related to this approach is the asymptotic iteration method, for which we refer the interested reader to references~\cite{Ciftci_2003,Ciftci:2005xn,Cho:2009cj,Cho:2011sf}.

\subsubsection*{Shooting Method}

The shooting method, also known as direct integration, is a completely numerical approach and well known in many other fields. 
The idea is that one explicitly integrates the perturbation equation from the boundaries for an initial guess for the QNM frequency and matches the numerical solutions at an intermediate point $x=x_m$ by computing the Wronskian
\begin{align}
W\left[\psi_1, \psi_2 \right](x, \omega) = \psi_1^\prime(x, \omega) \psi_2(x, \omega) - \psi_1(x, \omega) \psi^\prime_2(x, \omega).
\end{align}
If the chosen QNM frequency $\omega$ is an eigenvalue $\omega_n$ of the problem, the Wronskian vanishes because the solutions are linearly dependent. 
This allows one to find the spectrum of eigenvalues as solution of a root finding problem. 
The main advantage of the shooting method is that it is in principle rather straightforward to implement and can provide accurate results for small $n$. 
However, due to the strong overtone damping it becomes numerically very difficult to compute the spectrum for $n \gg 0$. 
The reason is that the radial functions towards the horizon and infinity correspond to exponentially diverging solutions, whose integration to an intermediate point where the Wronskian is evaluated is numerically unstable. 
The larger the damping of a given overtone, the less stable becomes the integration. 
To address this problem, the integration can start closer to the intermediate point, but this requires a very careful treatment of higher-order boundary conditions to achieve accurate results, which may make it cumbersome for some applications. 
In the context of black hole QNMs, it was first used by Chandrasekhar and Detweiler~\cite{Chandrasekhar:1975zza}, who transformed the Schr\"odinger wave equation to the Riccati equation and then applied the shooting method to it.

\subsubsection*{Time Evolution}\label{sec_methods_timedomain}

While all previously mentioned techniques provide the solution of the eigenvalue problem from the time independent perturbation equation, one can also obtain the spectrum from the explicit time evolution. 
This approach is arguably the most general one with respect to the previous methods, but also computationally more expensive and non-trivial in terms of extracting the QNMs. 
The time evolution of a Gaussian wavepackage in the Schwarzschild space-time was first studied by Vishveshwara in reference~\cite{Vishveshwara:1970zz} and the time evolution of a radially infalling test particle following geodesic motion in reference~\cite{Davis:1971gg}. 
In reference~\cite{Bachelot:1993dp} several modes have been extracted for the Schwarzschild black hole and in references~\cite{Krivan:1996da,Krivan:1997hc} for scalar and gravitational perturbations of the Kerr black hole. 
Due to the strong black hole overtone damping, the method is limited to small $n$. 
Note that for other types of compact objects, {\it e.g.}, neutron stars or some exotic compact objects, the direct time evolution can provide much better results than for black holes, because overtones are often much less damped and thus easier to extract. 
The advantage of the time evolution approach is that one can also understand the evolution of perturbations, which is particularly important when studying the relevance or significance of QNMs with respect to excitations, see reference~\cite{Nollert:1996rf} for a seminal work and section~\ref{sec3_relevance} for the discussion of this topic.

\subsubsection*{Numerical Relativity}\label{sec_methods_nr}

Another possible approach that we want to briefly mention is to solve the full Einstein field equations numerically which is commonly known as numerical relativity. 
Because there is a separate chapter in the full volume of this review dedicated to it, we only mention some basics here. 
Pioneering works in the mid 2000's~\cite{Pretorius:2005gq,Baker:2005vv,Campanelli:2005dd} and advances in large scale computer infrastructure make numerical relativity a rapidly growing and important field of research. 
While all previous methods are tools to study the effective wave equations of linearized perturbations around a static/stationary background, numerical relativity allows to study the full, non-linear Einstein field equations for suitable initial data. 
This is tremendously important in the study of binary black hole or neutron star mergers, whose merger phase cannot be captured by perturbative techniques or post-Newtonian calculations alone. 
The ringdown part can still be followed with numerical relativity, which allows one in principle to extract the QNM spectrum with similar techniques used in the previous paragraph on time evolution. 
Another great advantage is that it can provide the excitation amplitudes and phases for a given initial configuration, details of the initial binary black hole system's properties and hints of when perturbation theory becomes valid, see references~\cite{Buonanno:2006ui,Dorband:2006gg,Berti:2007dg,London:2014cma,Bhagwat:2017tkm,Giesler:2019uxc,JimenezForteza:2020cve,Cook:2020otn,MaganaZertuche:2021syq} for related studies on these aspects. 
Testing general relativity with a consistency of phases and amplitudes has been proposed in \cite{Forteza:2022tgq}.
While numerical relativity may sound like the most accurate and general approach, it also has several drawbacks that one should be aware of. 
It is computationally extremely heavy compared to all other previously outlined approaches, {\it i.e.},  simulations are non-trivial to setup and can easily take order of weeks on a computing cluster. 
Unfortunately, these simulations suffer from similar limitations as the ones for linear perturbations, which means one is limited to a few QNM overtone numbers $n$. 
We discuss some numerical relativity related approaches to theories beyond general relativity in section~\ref{sec5_overview_theories}. 

\begin{landscape}
\begin{table}
\begin{center}
\begin{tabular}{ c || c | c | c | c | c | c | c}
\textbf{Method} & Spectrum & Accuracy & Simplicity & Universality & Efficiency & ECO appl. & Literature\\
\hline \hline
\textbf{(Semi-)Analytic} &  & & & & & & \\ 
\hline \hline
Eikonal  & $n \lesssim l$ & low (mod. $\ell$) & high & medium & high & -- & \cite{1971ApJ...170L.105P,1972ApJ...172L..95G,Ferrari:1984zz,Ferrari:1984ozr}\\ 
\hline
(higher-order) WKB method & $n \lesssim l$ & high & high & medium & medium/high & no & \cite{Schutz:1985km,Iyer:1986np,Iyer:1986nq,Kokkotas:1988fm,Seidel:1989bp,Konoplya:2003ii,Matyjasek:2017psv,Konoplya:2019hlu}\\ 
\hline
Phase integral  & flexible & high & medium & medium & medium & -- &\cite{Froeman:1992gp,Andersson:1992scr,NAndersson_1993,Andersson:1995vi}\\  
\hline 
Analytic inverted potential  & $n \lesssim l$ & low & high & medium & high & -- &\cite{Mashhoon:1982im,BLOME1984231,Ferrari:1984ozr,Ferrari:1984zz,Churilova:2021nnc} \\ 
\hline
WKB inverted potential & $n \lesssim l$ & high & medium & medium & medium/high & no & \cite{Zaslavsky:1991ug,Hatsuda:2019eoj,Matyjasek:2019eeu} \\ 
\hline 
Parametrized QNM  & $n\leq2$ & medium & high & medium & high & no & \cite{Cardoso:2019mqo,McManus:2019ulj,Kimura:2020mrh,Volkel:2022aca} \\  
\hline
Monodromy & ``$n=\infty$'' & high & medium & medium & medium & -- &  \cite{Motl:2003cd,Cardoso:2004up,Keshet:2007be,Kao:2008sv} \\
\hline \hline 
\textbf{Full Numerical} &  &  & & & & & \\ 
\hline \hline
Shooting  & small $n$ & medium & medium & high & medium & yes & \cite{Chandrasekhar:1975zza} \\
\hline
Shooting inverted potential  & $n \lesssim l$ & medium & medium & high & medium & -- & \cite{Volkel:2022ewm} \\ 
\hline
Leaver & flexible  & high & medium & medium & high & -- & \cite{Leaver:1985ax,Rosa:2011my,Pani:2012bp,Pani:2013pma,Volkel:2022aca,Nollert:1993zz,Zhidenko:2006rs}  \\
\hline
Asymptotic iteration  & flexible  & high & medium & medium & high & -- & \cite{Ciftci_2003,Ciftci:2005xn,Cho:2009cj,Cho:2011sf} \\
\hline
Spectral methods & flexible & high & medium & medium & medium & -- & \cite{Jansen:2017oag,Fortuna:2020obg,Konoplya:2022zav,Chung:2023zdq} \\
\hline
Time evolution  & small $n$ & medium & medium & high & low & yes & \cite{Vishveshwara:1970zz} \\
\hline
Numerical relativity & small $n$ & medium & low & medium & low & yes & \cite{Buonanno:2006ui,London:2014cma} \\
\end{tabular}
\caption{
The purpose of this table is to give a rough overview of commonly used methods and to illustrate that each method has its own strengths and weaknesses. 
Rather than asking what is the best method, this table should provide some guidance about which method is more/less suitable for the particular application in mind. 
Of course this cannot take into account all details and subtle aspects of each method, which can be found in the provided references.  
We refer to \textbf{low}, \textbf{medium} and \textbf{high} as a \textbf{qualitative measure} with respect to most other methods in this table. 
Note that these simple ratings are to some extend subjective and not only reflect the authors personal opinion, but can also change depending on the specifics of the application in mind! 
Moreover, some methods are closely related to each other, which we further outline in the main text where applicable. 
We compare the different methods in the following categories. 
\textbf{Spectrum}: what part of the QNM spectrum can be computed?
\textbf{Accuracy}: how accurate is the method in its valid part of the spectrum? 
\textbf{Simplicity}: how easy is it to implement/use the method? 
\textbf{Universality}: can the method be extended to non-standard cases? 
\textbf{Efficiency}: how efficient is it to compute the QNMs? 
\textbf{ECO applicable}: has/can the standard method (without major adjustments/re-derivations) been used for exotic compact objects? 
\textbf{References}: seminal works and examples. 
In cases where we are not aware of existing literature or cannot confidently provide a rating, we indicate ``--''.
}
\label{table_qnm_methods}
\end{center}
\end{table}
\end{landscape}

\subsection{Relevance of Quasi-Normal modes}\label{sec3_relevance}

The methods presented in section~\ref{sec3_methods_qnms} allow to compute the QNM spectrum within perturbation theory, but another here yet unsolved and crucial aspect is to describe their relevance in the full binary merger context, not only their frequency and damping times. 
Therefore one must quantify the relevance of the QNMs in this bigger picture. 
Important, but non-trivial questions that arise are for example: what are the most dominant QNMs, how many modes are excited, how much does the excitation depend on the properties of the initial system? 

Besides special cases, {\it e.g.}, the test particle limit, one cannot completely answer most of the questions with perturbation theory alone. 
To be more precise, the excitation of a given QNM can be further split into a product of an excitation factor, which is independent of the initial data and is characteristic for the black hole, as well as an initial data/source dependent one. 
Numerical relativity computations, which evolve the late inspiral stage throughout the violent merger, can in principle be used to address these questions for any given system. 
Before we review this part we want to discuss some aspects that can be addressed within perturbation theory, or at least provide strong indications of some aspects that numerical relativity simulations should show. 

The relevance (or significance) of the QNM spectrum in the time domain has been highlighted in a simple scenario studied by Nollert in reference~\cite{Nollert:1996rf}. 
He considered different types of approximations of the RW potential in terms of square well potentials that mimic the exact potential when many small steps are used. 
In this approach the QNM spectrum can be computed as boundary value problem by matching (many) analytic solutions. 
Somehow surprisingly, one does not find good approximations to the RW QNMs, but very different spectra whose details depend on how many square potentials are being used and how they have been constructed. 
Earlier studies~\cite{Ching:1993gt,PhysRevA.49.3057} already revealed that introducing discontinuities in QNM systems can change the eigenvalue spectrum in a drastic way and make them a complete basis. 
Does this imply that small changes in the potential make the QNM spectrum irrelevant? 
In the same context, techniques like the inverted potential method \cite{Mashhoon:1982im,BLOME1984231,Ferrari:1984ozr,Ferrari:1984zz,Churilova:2021nnc} do provide meaningful results for the spectrum by studying approximate potentials. 
Whatever the modifications of the potential are, one may argue that the more important question to address is the evolution of initial perturbations.  
By studying the scattering of wavepackages, Nollert found that the time evolution is very similar to the one when using the exact RW potential, although traces of the true (non RW) spectrum become visible at late times. 
These findings have then been quantified in a more systematic study in reference~\cite{Nollert:1998ys}. 

Other situations in which the time evolution does not give the true QNM spectrum of a modified potential, but rather the expected result of the unperturbed case can be found from modified boundary conditions on the ``horizon scale'' of exotic compact objects (see. section~\ref{sec4_exotic_compact_objects}). 
The study of the stability of the QNM spectrum itself with respect to small perturbations of the potentials, {\it e.g.},  via environmental effects~\cite{Barausse:2014tra}, has received renewed interest in recent years. 
Even more recently the spectral stability has been revisited in terms of the so-called pseudospectrum method. 
We discuss both aspects in section~\ref{sec6_environmental}.

\subsection{Non-linear vs. linear Regime of the Ringdown}\label{sec3_linear_vs_nonlinear}

Before summarizing the key points about the possibility of non-linear features in the ringdown, we want to remind (from earlier discussions) the fact that even within linear perturbation theory, the ringdown is not fully characterized by a superposition of QNMs, because they do not form a complete basis. 
For a recent and very detailed review of the linear and non-linear ringdown analysis problem, we refer the interested reader to reference~\cite{Baibhav:2023clw}. 
In the following we will sketch some of the main aspects qualitatively. 

From time-domain calculations, {\it e.g.}, Gaussian wavepackages being scattered with the potential barrier, one knows that the early time response also contains non-QNM contributions whose form depends on the initial data. 
Very late times are dominated by a power law behavior known as the Price tail \cite{Price:1971fb,Price:1972pw}, which further depends on the asymptotic behavior of the potential at large distances. 
Thus, \textit{schematically} one can think of black hole perturbations that mimic the ringdown consisting of these parts
\begin{align}\label{sec6_lin_app}
h(t) = h_\mathrm{initial}(t) + h_\mathrm{QNM}(t) + h_\mathrm{tail}(t).
\end{align}
Thus even in linear perturbation theory, it is subtle to extract QNMs from the time-domain and one is usually restricted to the fundamental mode of a given harmonic $l,m$ and very limited with respect to overtones. 
By using different fitting models for waveforms obtained by two different perturbation potentials, reference~\cite{Nee:2023osy} recently stressed the difficulties in a standard overtone analysis, even in the linear regime. 

Let us now discuss the problem of non-linear effects. 
Binary merger simulations carried out with numerical relativity solve the full, non-linear Einstein equations. 
Hence, there are potentially non-linear contributions to the ringdown, which should become negligible at late enough times, where perturbation theory is accurate, but could be significant directly after the merger. 
One could think that the perturbation, similar to eq.~\eqref{sec6_lin_app}, can be written as 
\begin{align}
h(t) = h_\mathrm{non-linear}(t) + h_\mathrm{initial}(t) + h_\mathrm{QNM}(t) + h_\mathrm{tail}(t).
\end{align}
There are competing effects that make the analysis of linear and non-linear time-domain simulations difficult in practice. 
Because black hole QNMs have very short damping times, their extraction (in particular of the overtones) may depend on when exactly the modes are being extracted. 
The crucial aspect is the \textit{starting time} of when the ringdown can be adequately described in terms of a superposition of damped modes in linear theory!

If the starting time of the ringdown analysis is too early, with respect to the non-linear merger part or initial data, one may include rapidly damped overtones, but may also bias their values by including non-linear parts that do not truly correspond to the linear modes one is looking for. 
Waiting too long on the other hand makes sure that the non-linear parts are negligible, but also implies that the overtones are already damped so much that the fundamental mode dominates and the overtones are already too weak to be identified. 
Moreover, the very late time behavior of the ringdown is not described by the fundamental QNM but the power law tail which further restricts one from choosing very late starting times. 

How do non-linear features look like? In general this is hard to tell, but one could gain insights from second order perturbation theory. 
Instead of carrying out the more involved computations, see reference~\cite{Nakano:2007cj,Lagos:2022otp} for approaches for the Schwarzschild case and reference~\cite{Ripley:2020xby} for Kerr, one might also model them in a phenomenological way by assuming that linear modes couple to each other. 
This predicts a new set of ``second order QNMs'' that are determined from combinations of the linear ones. 
In reference~\cite{London:2014cma} such an ansatz has been suggested and used to analyze numerical relativity simulations. 
More recently in references~\cite{Cheung:2022rbm,Mitman:2022qdl} similar analyses have been carried out, and in references~\cite{Dhani:2020nik,Dhani:2021vac} related studies including mirror modes. 
Together with results from references~\cite{Sberna:2021eui,Baibhav:2023clw} one should thus take non-linear effects in the ringdown seriously. 
This evidence for non-linearities is somewhat in contrast with the results obtained in reference~\cite{Giesler:2019uxc}, where it is argued that by using a large number of ordinary overtones one can well describe binary mergers beyond the peak. 
It is argued that this would not correspond to overfitting, because other tests have been performed, {\it e.g.}, it is also shown that the final mass can be estimated more accurately at earlier times when the overtones are being included. 
This finding might to some extend also be understood from the excitation factors of the Kerr black hole, which seems to suggest that the first few overtones may indeed be excited significantly in the ringdown~reference~\cite{Oshita:2021iyn}. 
A possible tool to model the non-linear ringdown for Kerr, partially inspired by related approaches for optical cavities~\cite{PhysRevA.49.3057}, has been suggested in reference~\cite{Green:2022htq} and consists of a product under which QNMs are orthogonal.

\begin{svgraybox}
In this section we discussed:
\begin{itemize}
\item how the QNM spectrum can be computed with commonly used methods;
\item what the significance of the QNM spectrum is in terms of time evolution;
\item possible limitations of linear perturbation theory in the ringdown and problems of overtone fitting;
\item that current literature is not providing a complete answer to understanding the ringdown.
\end{itemize}
\end{svgraybox}

\section{Possible Deviations from General Relativity}\label{sec5}

In this section we discuss the diverse landscape of how general relativity, and in particular QNMs, could be modified in alternative theories of gravity or exotic compact objects. 
We review theory specific aspects in section~\ref{sec5_overview_theories}, theory agnostic and phenomenological approaches in section~\ref{sec4_agnostic_pheno}, outline exotic compact objects and echoes in section~\ref{sec4_exotic_compact_objects}, and discuss the inverse problem in section~\ref{sec4_inverse}.

\subsection{Overview QNMs in Modified Gravity Theories}\label{sec5_overview_theories}

From the frequency domain analysis of the Teukolsky equation, it appears clear that the only free parameters appearing in equations~\eqref{eq:SpheroidalHarmonics}--\eqref{eq:Teukolsky_Radial} are the mass $M$ and the spin $a$ of the black hole. This property is a consequence of what is commonly known the {\it no-hair conjecture}, which states that stationary vacuum solutions with asymptotically flat boundary conditions in general relativity are described uniquely by their mass and spin (and possibly electric charge)~\cite{Israel:1967wq,Carter:1971zc,Robinson:1975bv}. By assuming certain conditions on the metric, one can mathematically prove the absence of additional degrees of freedom around black holes. These results are known as {\it no-hair theorems}, where additional information about them can be found in specialized reviews~\cite{Herdeiro:2015waa,Sotiriou:2015pka}.

The fact that no-hair theorems exist means that a violation of some of the hypotheses could lead to a so-called hairy black hole. It is convenient to split the discussion in two separate cases when no-hair theorems are circumvented: general relativity with exotic matter and beyond general relativity theories. In the first case we collect all of those cases which do not modify the dynamical properties of the gravitational field, but somehow encompass some non-standard matter content in the theory. Examples can be the inclusion of an electromagnetic field which leads to the Kerr-Newman solution, or a massive scalar/vector field that can trigger superradiant instability, and if complex, lead to stationary long-lived scalar cloud configurations around the black hole. On the other hand, one can have non-trivial black holes by modifying the underlying theory of gravity. 

In all of the cases discussed, it is expected that the spectrum of QNMs would be different from that of Kerr black holes, presented in section~\ref{sec2}, for some of the following reasons
\begin{itemize}
    \item The background metric is different;
    \item The dynamics of the theory is different;
    \item Couplings with additional fields enrich and modify the spectrum;
    \item Boundary conditions can vary.
\end{itemize}
In general, these features affect the calculation of the perturbation equations and of the QNMs in different ways for different theories. In the following, we try to summarize the main trends for some selected theories, whose behaviour covers one or more of the aforementioned bullet points as well as the strategies proposed to solve the problem. This non-exhaustive selection was made on the base of: availability of results, presence of hairy black holes in the theory, possible problems that arose in the computation, viability of the theory according to current tests.

\subsubsection*{Non-rotating case}

The equations governing perturbations were obtained in many beyond general relativity cases, at least in the non-rotating limit. In general though, it is not obvious how to reduce them to a Schr\"odinger-like form as for the RW and the Zerilli equations~\eqref{eq:RW_Zerilli}. An illustrative case of why this is not always possible is analysed in~\cite{Langlois:2021aji}. Anyway, this property does not hinder from computing the spectrum of QNMs of the theory just by solving all the perturbation equations numerically, although one has to be cautious about the implementation of boundary conditions~\cite{Langlois:2021xzq}. This approach can be seen as a generalization of the direct integration method described in section~\ref{sec3_methods_qnms}, and the applications, apart for the aforementioned Horndeski theories, can be found for the more general degenerate-higher-order scalar-tensor (DHOST) theories~\cite{Langlois:2022ulw} or for the special case of scalar-Gauss-Bonnet (sGB) gravity~\cite{Blazquez-Salcedo:2016enn,Blazquez-Salcedo:2017txk,Blazquez-Salcedo:2020rhf,Blazquez-Salcedo:2020caw,Langlois:2022eta}.

We have seen that having a Schr\"odinger-like form for the perturbation equation is not a necessary requirement for the computation of the QNM spectrum, but there are cases in which it is desirable to have it, for the application of precise methods of the computation (such as the continued fraction one). A very useful way to obtain this reduction is by performing a small-coupling (SC) expansion. With SC we refer to a perturbative expansion in the coupling $\al$ for any theory whose action can be thought of the form
\begin{equation}
    S_{tot}[g_{ab},\Phi_i] = S_{GR}[g_{ab}] + S_{\Phi}[g_{ab},\Phi_i] + \al S_{int}[g_{ab},\Phi_i].
\end{equation}
Here, the action is intended as a modification of the Einstein-Hilbert action $S_{GR}$ where $g_{ab}$ is the spacetime metric and $\Phi_i$ are all the additional fields whose free dynamics is described by the action $S_\Phi$ and $S_{int}$ describes the non-minimal interaction between gravity and new fields. For example, if there is only one field $\Phi_i$ and it is a scalar field $\phi$, then $S_\Phi$ is the action given in equation~\eqref{eq:scalar_action}, and $S_{int}$ would depend on the theory in consideration. The SC limit is taken when $\al \ll 1$ and it drastically simplify the equations in many cases. The motivation behind this approach is that current detections of GWs are within percent agreement with general relativity, meaning that drastic modifications are likely to be disregarded. 

The SC limit allows the application of two very useful methods to compute deviations of the frequencies given a small deviation from the general relativity potential: parametrized deviations and perturbative frequency expansion. The first method, available only for the spherically symmetric setup, consists in linking the deviations appearing in the potential to deviations in the frequencies, through a Taylor expansion. It was developed in~\cite{Cardoso:2019mqo}, extended to second order in the SC and multiple field case in~\cite{McManus:2019ulj} and more accurate coefficients, up to second overtone number for all the cases previously considered in~\cite{Volkel:2022aca}. The advantage of this method is its universality, in the sense that the deviation coefficients for the frequencies do not depend on the theory considered, which only enters through the coupling $\al$. The main disadvantage is that it is a scheme based on strict requirements of the shape of the perturbation equation, which makes it suitable only for specific theories.

The perturbative frequency (PF) expansion method is inspired from quantum mechanics and it was first applied for QNM computation in~\cite{Mark:2014aja}. The idea is inspired from quantum mechanics, for the case of study of shifts in energy spectra time-independent Schrodinger equation. The method relies on the definition of a self-adjoint scalar product onto which project the equation and extract perturbatively the modification to the frequency. It is a rather general method that fits well within the SC limit.

\subsubsection*{Rotating case}

For the spinning case, things are more complicated than in spherical symmetry. In fact, the two main issues are that we do not know analytic closed form solutions for rotating black holes in the majority of theories, and the perturbation equations are not separable, as in the Teukolsky approach shown in section~\ref{sec:Teukolsky}. There are different ways to tackle these two problems, either by analysing the system in a specific limit, or by solving numerically the whole system of equations. 

Among the limits that turned out to simplify the problem we can find the eikonal limit, taken when $\ell\to\infty$ and that we already encountered in section~\ref{sec3_methods_qnms}. Despite this method is rather straightforward in the application, it provides only a qualitative behaviour of the QNM spectrum of a given theory, especially for low angular momentum modes ($\ell=2,3$) which are the most dominant ones in the ringdown signal. The advantageous feature of the eikonal approximation is that it is easily applicable to a vast class of theories without drastic modifications in the method~\cite{Silva:2019scu,Dey:2022pmv}.

\begin{figure}
\centering
\includegraphics[width=0.7\linewidth]{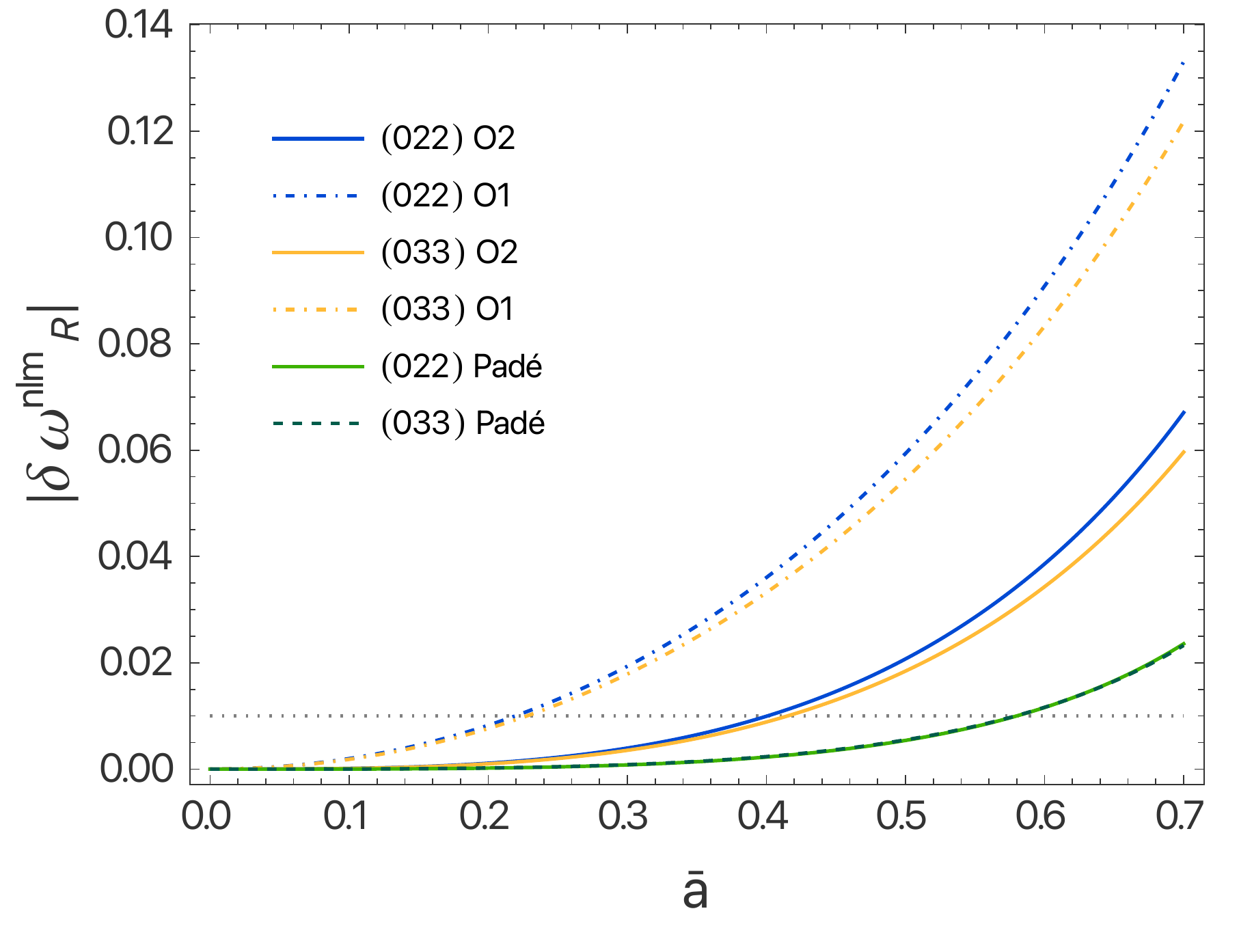}
\includegraphics[width=0.7\linewidth]{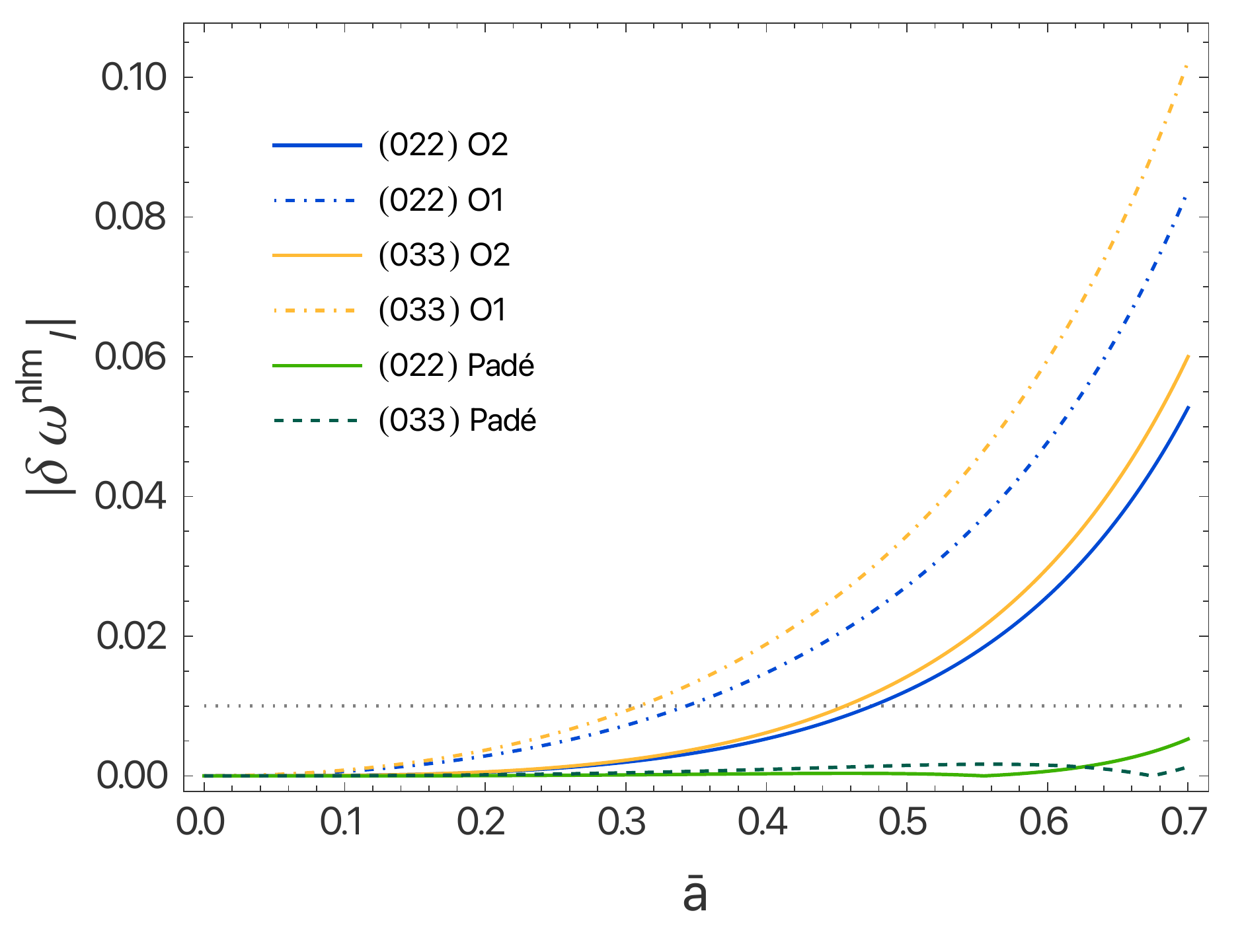}
\caption{ Real (upper panel) and imaginary (lower panel) parts
of the relative difference between the QNMs of Kerr black holes and
those of rotating black holes computed within the slow-rotation approximation, for the $(nlm) = (022),\,(033)$ modes. The SR expansion is performed to first order (O1), to second
order (O2), to second order with Pad\'e resummation (Pad\'e).
The horizontal dotted line represents a 1\% error. Credit:~L. Pierini and L. Gualtieri, Phys. Rev. D 106 (2022) 104009.
doi:10.1103/PhysRevD.106.104009 \cite{Pierini:2022eim}.}
\label{fig:SR_expansion}
\end{figure}

A more quantitative approach is the slow-rotation (SR) or slow-spin approximation. As suggested by the name, on top of the perturbation scheme one performs an additional expansion in the spin, up to desired order. This idea was first developed by Kojima in the early 90s in the contest of oscillation of slowly rotating stars~\cite{Kojima:1992ie}. The first application to the context of black holes physics was for the stability analysis of massive {\it spin}-1 fields around Kerr black holes~\cite{Pani:2012vp,Pani:2012bp}, and it was only recently that QNMs spectra were obtained for some theories up to first order: higher derivative gravity~\cite{Cano:2021myl}, sGB~\cite{Pierini:2021jxd} and dynamical Chern-Simons (dCS)~\cite{Wagle:2021tam,Srivastava:2021imr}; and up to second order for sGB~\cite{Pierini:2022eim}. One can think that the drawback of SR expansion is that it does not give information about spectrum of medium-high spinning black holes, which are those relevant for observations. However, in~\cite{Pierini:2022eim}, the authors showed that with a diagonal Pad\'e resummation~\cite{bender1999advanced}, they manage to obtain reasonable results up to $a \sim 0.7$. In Fig~\ref{fig:SR_expansion} we report their results on the comparison of the Kerr modes computed in the SR expansion truncated at first order, second order and second order plus Pad\'e resummation. One can clearly see from the plot the improved performance of the resummation, for which relative errors fall well below the 1\% tolerance line even at spins $a\sim0.6$ for the real part and $a>0.7$ for the imaginary part. This seems, up to date, the most promising method to compute useful QNMs for tests of general relativity.

Recently, it was pointed out that SC expansion can be useful even for the study of perturbations in the rotating case. Cano, Fransen and Hertog analysed this possibility for a scalar perturbation on top of a rotating black hole in higher-derivative gravity~\cite{Cano:2020cao}. The main result of that study is a prescription to separate coupled system by a suitable projection of the equations onto spheroidal harmonics, which was later generalised to any SC system in~\cite{Ghosh:2023etd}. The situation for the full gravitational case is much more complex, since one would need to generalize the Teukolsky equation for generic theories of gravity~\cite{Li:2022pcy,Hussain:2022ins}. Up to now, there is only one attempt of applying the SC limit for rotating black holes in higher-derivative gravity~\cite{Cano:2023tmv}. The fact that it does not bind the analysis to small spins makes it a promising line for future research.

Along the line of expanding the system of perturbations either in angular momentum, spin or coupling of the theory, one might want to be able to compare these results with full-theory results. In the most general case, one would have up to 10 coupled partial differential equations, for which the only reasonable way to solve the problem is by numerical integration. Up to now, this approach was used to cover the computation of QNMs for Kerr-Newman black holes~\cite{Dias:2021yju,Dias:2022oqm}, for which the symmetries of the solution reduce the perturbation system to two coupled partial differential equations. It was proposed that spectral methods are a good numerical approach for the solution of the full ten equations system, and it was applied to Schwarzschild case only~\cite{Chung:2023zdq}.

Finally, we want to conclude by commenting on the possibility of using time-domain numerical relativity simulation to infer QNMs of alternative theories of gravity. This would in principle be doable, but it could be employed only once the current status of most alternative theories is resolved. We are referring to the formulation of the initial data problem {\it \'a la} Cauchy, which is well-posed for some theories, but this is not always the case. For a deeper discussion on the topic and possible solutions to this issue we remand to the chapter of this book on numerical relativity. It is clear that the current status of numerical relativity beyond general relativity is still under development, and perturbations were analysed only for a handful of cases either to asses stability of the solution~\cite{Okounkova:2018pql,Okounkova:2019zep}, or scalar QNMs in decoupling limit~\cite{Evstafyeva:2022rve}.

We conclude this section by summarizing the information into two tables. In the first table~\ref{tab:spin_approximations}, we collect all the methods that have been proposed to simplify or tackle the study of QNMs of rotating black holes in alternative theories of gravity, stating their current status, accuracy reached at spin $a\sim0.7$, and a qualitative comment about simplicity of the application of the model and how it can universally extended to other cases.
In the second table~\ref{tab:beyondGR}, we report the status for some selected theories, comparing results for non-rotating and rotating case, and stating what is the current most precise result, what is the highest overtone, angular momentum and spin reached and which method was used to obtain it.
\begin{table}
\begin{center}
\begin{tabular}{ c || c | c | c | c | c}
{\bf Method} & Status & Accuracy at $a=0.7$ & Simplicity & Universality & References \\
\hline \hline
Eikonal  & Completed & Depend on $\ell$ & High & High & \cite{Silva:2019scu,Dey:2022pmv} \\
SR  & Up to 2nd order & Pad\'e resummation & Medium & Medium & \cite{Cano:2021myl,Pierini:2022eim,Wagle:2021tam,Srivastava:2021imr} \\
SC  & Up to 1st order & Not known & Low & Medium & \cite{Li:2022pcy,Hussain:2022ins,Cano:2023tmv} \\
FDI & Kerr-Newman & Theory-dependent & Very low & No & \cite{Dias:2021yju,Dias:2022oqm,Chung:2023zdq} \\
NR & / & / & Very low & No &
\end{tabular}
\caption{Summary of the adopted techniques to simplify the problem of computation of QNMs for rotating black holes beyond general relativity. For each method (SR = slow rotation, SC = small coupling, FDI = frequency domain integration, NR = numerical relativity), we report the current status of completion, plus a comment about status of accuracy of the calculation for $a\sim0.7$, simplicity and universality of the implementation of the model and some key literature.}
\label{tab:spin_approximations}
\end{center}
\end{table}

\begin{table}
    \centering
    \begin{tabular}{cc||cccc|ccccc}
    \multirow{2}*{\bf Theory} & \multirow{2}*{Coupling} & \multicolumn{4}{c|}{Non-rotating} & \multicolumn{5}{c}{Rotating} \\
         & & $n$ & $\ell$ & method & reference & $a_{max}$ & $n$ & $\ell$ & method & reference \\
         \hline \hline
       GR + EM & $Q$ & $10^5$ & any & CF & \cite{Leaver:1990zz,Berti:2003zu} & $\sqrt{1-Q^2}$ & 1 & 2 & FDI & \cite{Dias:2021yju,Dias:2022oqm} \\
       Higher-derivative &  & 0 & 4 & DI, SC & \cite{Cardoso:2018ptl} & 0.4 & 0 & 3 & DI, SR/SC & \cite{Cano:2023tmv} \\
       \multirow{2}*{sGB} & $f'(\phi) \neq 0$ & 0 & 3 & DI & \cite{Blazquez-Salcedo:2016enn,Blazquez-Salcedo:2017txk} & 0.7 & 0 & 3 & DI, SR & \cite{Pierini:2022eim} \\
        & $f'(\phi) = 0$ & 0 & 2 & DI & \cite{Blazquez-Salcedo:2020caw}
        & & & & & \\
       dCS & & 2 & 4 & CF, SC & \cite{Volkel:2022aca} & 0.2 & 0 & 4 & DI, SR & \cite{Wagle:2021tam,Srivastava:2021imr} \\
       Horndeski &  & & & & \cite{Tattersall:2017erk,Langlois:2021xzq} & & & & & 
    \end{tabular}
    \caption{For each black hole solutions in a given theory, we report the most up-to-date result on QNMs for non-rotating solutions and rotating solution. For each case, we report the coupling if it has some relevance, the maximum overtone number $n$, the maximum angular momentum number $l$ of the spectrum, as well as the most precise method used for the calculation (CF = continued fraction, DI = direct integration, FDI = frequency domain integration) and, if present, the prescription that leads to some simplification of the problem (SC = small coupling, SR = small rotation). For the spinning case we report also the highest value of the spin $a$ for which the results can be trusted}
    \label{tab:beyondGR}
\end{table}

\subsection{Theory-Agnostic and Phenomenological Approaches}\label{sec4_agnostic_pheno}

Along the research into the QNM spectra of black holes in alternative theories of gravity, considerable effort is put into the development of ringdown models that are not based on a specific theory, but rather try to reproduce as many features as possible with a bunch of free parameters. This bottom-up approach is particularly suitable for comparison with real data, as ideally one would have few parameters that can be constrained and eventually traced back to the cause of the modification. We already encountered a model which is suitable for this purpose: the parametrized framework developed in~\cite{Cardoso:2019mqo,McManus:2019ulj} offers a direct way to describe how single modifications in the effective potential globally affect the spectrum of frequencies. The main obstacle to this method for real data comparison is that so far it was developed only for the non-rotating limit. Clearly, one can apply it coupled with a SR expansion, where again diagonal Pad\'e resummation seems to suggest that high values for the spin can be achieved~\cite{Hatsuda:2020egs}.

Despite the lack of intuition of what happens to perturbations of rotating black holes within most theories, Maselli {\it et al.~}tried to derive a phenomenological model for modifications of spectra of spinning black holes dubbed {\it ParSpec}~\cite{Maselli:2019mjd}. This model is based on the assumptions that if a modification is present, then it would be small and it can be fitted by a low order polynomial function in the spin. Then, three separate cases can appear: the modification can either come from some coupling constant, dimensionful or dimensionless, or from a free parameter that describes the black hole. In this way, one can summarize with a relatively small number of parameters the information of how any modification is originated.

The main issue with any parametrized formalism is that it is hard to connect them with the fundamental principles that are behind a possible modification in the spectrum. For this reasons, there have been attempts in formulating an {\it effective field theory} of QNMs, that parametrizes modifications in the frequencies directly from a handful of parameters of some general enough action. The first attempt was made in~\cite{Tattersall:2017erk}, where the authors listed all the possible covariant terms that can affect perturbations on top of Schwarzschild black holes, including contributions from scalar and vector fields. A more general result, that encompasses any spherically symmetric black hole plus a scalar field coupled to gravity~\cite{Franciolini:2018uyq} and then extended to the slowly-rotating case~\cite{Hui:2021cpm}. The assumption made is of a {\it spin}-2, unitary, Lorentz invariant theory of gravity coupled to a {\it spin}-0 field, but in principle, one could add the operators that would appear if any of these hypotheses is not met. An effective field theory of QNMs of rotating black holes is still lacking.

One other approach that has been widely explored in the literature is to consider parametrized black hole metrics, which are describing hypothetical deviations from the Schwarzschild/Kerr metrics~\cite{Vigeland:2011ji,Johannsen:2011dh,Johannsen:2013szh,Rezzolla:2014mua,Konoplya:2016jvv,Papadopoulos:2018nvd}. 
They can be very useful for practical purposes, {\it e.g.},  when black hole metrics in modified theories are only known numerically in order to find a good approximation for further calculations like ray tracing~\cite{Younsi:2016azx}. 
However, because those metrics are not solutions to the vacuum Einstein field equations, such space-times can be pathological and any objective that requires one to know the underlying field equations, like computing QNMs, must be understood with caution. 
On the one hand, it is straightforward to study test field perturbations for a given metric. 
Since one knows that they are qualitatively similar to the full gravitational ones in general relativity, some smoking gun features could be found, at least qualitatively. 
Moreover, if one assumes that modifications to general relativity are small, one may also use the general relativity field equations as approximation to derive gravitational perturbation equations for such metrics~\cite{Barausse:2014tra,Volkel:2020daa,Franchini:2022axs}. 
If one assumes that the eikonal limit also holds for the underlying, unknown theory of a given black hole, it can be used as rough estimate to approximate the QNMs, see references in section~\ref{methods_eikonal}. 
On the other hand, there are more involved mechanisms, as mentioned in section~\ref{sec5_overview_theories}, so test field studies can hardly embody a complete analysis. 
Furthermore, without a direct connection to an underlying theory, {\it e.g.},  the action of coupling constants, any constraints are to some extend limited to phenomenological aspects, rather than theory specific ones. 
In reference~\cite{Suvorov:2021amy} this issue was somehow addressed by starting from the Rezzolla-Zhidenko metric~\cite{Rezzolla:2014mua}, then constructing a suitable scalar tensor gravity theory for which this metric is a solution, and only then study consistently the axial perturbation equations of that theory. 
The construction of theories for a given metric is not uniquely possible, and it was shown that for some theories using the general relativity background equations (as assumed in reference~\cite{Volkel:2020daa}) as approximation holds for some of them, but not all.

\subsection{Exotic Compact Objects and Echoes}\label{sec4_exotic_compact_objects}

In standard general relativity and ordinary nuclear physics there are only black holes and neutron stars as the two types of compact objects. 
Compared to vacuum black holes, neutron star physics is incredibly rich and there are many open problems, {\it e.g.},  finding the correct nuclear equation of state at high densities. 
However, in terms of their gravitational wave emission from a ringdown analysis, it is well understood that the characteristic modes are very different, even if both objects have a similar mass. 
This does not mean that it is always easy to discriminate between them in binary mergers. 
In the absence of an electromagnetic counterpart and low signal-to-noise ratio (SNR), the late inspiral and merger can look very similar and the characteristic modes may not be measurable. 
However, in the context of this review, we do not want to discuss neutron stars further and refer to references~\cite{Kokkotas:1999bd,Nollert:1999ji} for classical reviews on their oscillations and other chapters of this topical review. 
In the following we consider objects that are commonly known as ``ultra compact objects'', ``exotic compact objects'', ``horizonless compact objects'' or ``black hole mimickers''. 
We want to stress that standard black holes and neutron stars are widely accepted, there are many observations and theoretical considerations that make them viable objects in an astrophysical context. 
However, rather from the theoretical side, there are problems related to black holes that indicate a certain level of incompleteness of general relativity to describe everything about black holes when it comes to singularities and their relation to quantum mechanics. 
These problems are worth a review for themselves and it is impossible to cover all of them in a proper way. 
A more refined overview and classification of exotic objects in the gravitational wave context can be found in references~\cite{Cardoso:2017cqb,Cardoso:2019rvt}. 

The first related computation of the QNM spectrum of such objects had been carried out by Chandrasekhar and Ferrari in reference~\cite{doi:10.1098/rspa.1991.0104} for constant density stars approaching the Buchdahl limit and followed up in reference~\cite{Kokkotas:1994an}. 
The time evolution waveform that shows the characteristic ``echoes'' form of the late time response of the same system has first been demonstrated by Kokkotas in reference~\cite{Kokkotas:1995av} and was elaborated in subsequent works \cite{Tominaga:1999iy,Ferrari:2000sr}. 
These works have set the foundations of the so-called quasi-bound modes or trapped modes, which are the new type of QNMs that appear for ultra compact horizonless systems and the phenomenology of echoes. 
The key mechanism is that initial radiation that would have been absorbed by the horizon gets reflected, and then keeps bouncing between the potential barrier and the object. 
It gets partially transmitted through the potential barrier, which then gives rise to echoes for an observer far away, see Fig.~4 in reference~\cite{Kokkotas:1995av}. 
Among the most popular studied models are constant density and other ultra compact stars~\cite{doi:10.1098/rspa.1991.0104,Kokkotas:1994an,Kokkotas:1995av,Urbano:2018nrs}, anisotropic stars~\cite{Raposo:2018rjn}, boson stars~\cite{Kaup:1968zz,Liebling:2012fv}, wormholes~\cite{Damour:2007ap,Konoplya:2016hmd}, gravastars~\cite{Mazur:2001fv,Mazur:2004fk,Mottola:2023jxl}, fuzzballs~\cite{Ikeda:2021uvc}, area quantization~\cite{Bekenstein:1974jk,Mukhanov:1986me,Hod:2015qfc,Foit:2016uxn,Cardoso:2019apo,Coates:2019bun,Agullo:2020hxe,Coates:2021dlg} and collapsed polymers~\cite{Brustein:2016msz,Brustein:2017koc}. 
One important motivation to study alternatives to standard black holes is the information loss problem that might give rise to firewalls and quantum effects on the horizon scale. 
In the absence of a full theory which could be solved on the same footing as general relativity, such hypothetical modifications on the horizon scale are usually studied in a rather phenomenological way. 
A very popular approach to capture possible smoking gun signals is to replace the fully absorbing boundary condition at the horizon with some reflection very close to the would be horizon, {\it e.g.}, via membrane paradigm inspired models~\cite{Barcelo:2017lnx,Oshita:2019sat,Maggio:2020jml,Chakraborty:2022zlq}. 
The idea is that the true object is not a normal black hole but admits new physics that might influence gravitational waves. 
Note that this may not provide a full description of what this object is made of or whether it is stable. 
One possible instability mechanism that needs to be avoided is the so-called ergoregion instability, which is a quite common feature once rotation is being considered, see, {\it{e.g.}}, references~\cite{Cardoso:2014sna,Maggio:2017ivp} as examples. 
Models of exotic compact objects may rather be considered as an effective way to understand possible effects and put constraints on such deviations, see, {\it e.g.},  references~\cite{Mark:2017dnq,Correia:2018apm} for examples of approaches to construct echo waveforms. 
We provide a discussion on observational status of echoes and more recent attempts to model them in section~\ref{sec5_echo}.

\subsection{Inverse Problem for Quasi-Normal Modes}\label{sec4_inverse}

So far we have focused on what one could call the {\it direct problem}, in which one is interested in obtaining the QNMs for a given system. 
A complementary aspect is the so-called {\it inverse problem}, which assumes that a certain number of QNMs can be provided (from theory or observations) and one is interested to reconstruct the properties of the object. 
The most famous example related to spectral analysis is mostly attributed to Mark Kac with his work titled: \textit{Can one hear the shape of a drum?}~\cite{Kac:1966xd}. 
More specifically, given the entire spectrum of eigenvalues of an idealized drum, can one determine its shape? 
In that case the answer was provided in reference~\cite{1992InMat.110....1G} and is that the reconstruction is possible, but not unique. 
There are infinitely many shapes, obeying certain geometrical properties, that admit exactly the same spectrum. 
This non-uniqueness is quite common in inverse problems and makes them in general much more complicated and often ill-posed. 

For black holes the main question is whether one assumes that the underlying object is described by the Kerr metric within general relativity, or whether one considers an exotic compact object or/and theories beyond general relativity. 
The former one will in principle translate the observed QNMs to the mass and spin of the Kerr black hole. 
Although this is not trivial in practice, see for instance our discussion in section~\ref{sec5_multiple_qnms}, it ``only'' corresponds to fitting (inferring) the parameters of a very specific model (Kerr black hole) in a specific theory (general relativity) to observed data (QNMs). 

A different approach is to assume a minimal structure on the level of the perturbation equations, which one would expect for a range of different objects or theories, and then reconstruct the main properties. 
We have already discussed several related aspects in section~\ref{sec4_agnostic_pheno}. 
For non-rotating objects and in some beyond general relativity theories one expects the main perturbation equation to be of the time-independent Schr\"odinger equation with a potential term, see {\it e.g.}, reference~\cite{Cardoso:2019mqo}. 
The reconstruction of the potential in terms of theoretically provided QNMs can also be seen as the inversion of an operator and there is a large amount of literature from other fields of physics and mathematical methods, see reference~\cite{MR985100} for a standard textbook. 
The inverse problem has been explicitly studied for different classes of exotic compact objects in references~\cite{Volkel:2017kfj,Volkel:2018hwb} using WKB theory by combining results known from the ``inversion'' of the classical Bohr-Sommerfeld rule and the Gamow formula~\cite{Gamow:1928zz}, which were reported in references~\cite{lieb2015studies,Cole:1978zz}\footnote{Also see Chapter 4 in \cite{Volkel:2020lcy} for a comprehensive overview and more details on the application of WKB theory to the inverse problem.}. 
One main criteria in the question of the reconstruction is the number of classical turning points of the potential, which cannot be uniquely reconstructed from the QNMs alone. 
Moreover, one of the main observations is that also the potentials can in general not be uniquely reconstructed, but rather the separation of the classical turning points as function of the energy. 
However, by assuming that the assumptions of Birkhoff's theorem hold and that the potential outside the object can be mapped to the one of the Schwarzschild black hole, one can find approximate potentials valid under the assumptions of the used WKB methods. 
For more details on WKB equivalent potentials see references~\cite{Bonatsos_1991,Bonatsos:1992qq,Bonatsos:1992he}, and reference~\cite{Volkel:2018czg} for some ill-posed examples in the application of the WKB inversion methods. 

The alternative approach to address the inverse problem tailored to classical black holes by means of statistical inference has in recent years been applied to hypothetical data of QNMs~\cite{Volkel:2020daa,Volkel:2022aca,Volkel:2022khh} and to actual data~\cite{Dey:2022pmv}. 

\begin{svgraybox}
In this section we discussed: 
\begin{itemize}
\item open problems and different approaches to tackle them (table~\ref{tab:spin_approximations});
\item current status of the computations of QNMs in alternative theories of gravity (table~\ref{tab:beyondGR});
\item summary of the approaches for phenomenological description of QNMs;
\item the smoking gun effects in the late time signal of binary mergers in the forms of echoes;
\item that a given QNM spectrum can be used to infer properties of the object in the inverse problem.
\end{itemize}
\end{svgraybox}

\section{Confrontation between Theory and Observations}\label{sec6}

Equipped with the theoretical understanding of the previous sections, we finally come to the observational status of testing general relativity with QNMs. 
As this review is focusing on QNMs and not on gravitational wave physics in general, we will not discuss any details related to the measurement of gravitational waves and refer the interested reader to references~\cite{Maggiore:2007ulw,LIGOScientific:2016wyt} and proceed with the bare minimum. 
One may describe the current gravitational wave detectors of LIGO, Virgo and KAGRA (LVK) as large scale laser interferometers (a few km long arm-lengths) that allow to measure tiny variations of space-time in a frequency range that is most sensitive from a few Hz to a few hundred Hz. 
Despite advanced high technology gained over several decades, {\it e.g.},  by smaller detectors like GEO600, even viable signals are dominated by noise and require a careful treatment. 
This is taken into account by Bayesian statistical methods that allow to provide probability distributions for the parameter space of the initial binary system and final black hole, given an observed signal, correct noise modeling and a model to compute the underlying physical waveforms. 
The typical mass range of compact binary merger systems observable with current gravitational wave detectors ranges from a few solar masses to several tens of solar masses. 

After reviewing the detection of (multiple) QNMs in section~\ref{sec5_multiple_qnms}, we explain the differences between time-domain and frequency-domain tests in section~\ref{sec5_TDvsFD}, we discuss bounds on modified theories in section~\ref{sec5_theory_bounds} and report results on echo searches in section~\ref{sec5_echo}.

\subsection{Detecting (Multiple) Quasi-Normal Modes}\label{sec5_multiple_qnms}

The current status on the actual detection of QNMs from compact binary mergers observed by the LVK Collaboration is exciting, especially with respect to the possibility of having observed more than one in some events. 
So far it has been possible to robustly infer the $\ell=m=2$ fundamental mode frequency and damping time from the ringdown, most prominently for the first event GW150914~\cite{LIGOScientific:2016lio}, but also for other events~\cite{LIGOScientific:2020tif,LIGOScientific:2021sio}. 
The fact that the $\ell=m=2$ fundamental mode is the most common mode is clearly expected from perturbation theory and numerical relativity simulations. 
The fundamental mode and damping time extracted from the ringdown are in agreement with those of a Kerr black hole with a certain mass and spin, but this alone is not sufficient to test the Kerr metric. 
However, due to the independent extraction of mass and spin from the inspiral-merger part, it has been possible to verify that the extracted fundamental mode is consistent \cite{LIGOScientific:2016lio,LIGOScientific:2020tif,LIGOScientific:2021sio}. 
To truly perform black hole spectroscopy and unleash its full potential to probe the Kerr metric, it is necessary to observe at least a second mode (at least its frequency or damping time). 
Because the spectrum of all possibly excited QNMs is uniquely determined by mass and spin, any observed mode outside the prediction would be a clear violation. 

Unfortunately, there are several aspects that complicate black hole spectroscopy. 
In section~\ref{sec3_relevance} we discussed the relevance of QNMs to describe the time evolution of perturbations (in linear theory). 
While the QNM spectrum can be uniquely predicted, it is much more difficult to assess the excitation amplitudes that determine to what extend the modes are being excited, {\it e.g.},  which $(\ell, m, n)$ is sub-dominant and how many QNMs need to be included to avoid bias. 
Moreover, due to the incompleteness of QNMs, there is an early response that is a non-QNM contribution and thus complicates the analysis. 
Also, because the excitation comes from the non-linear merger of compact objects, this can only be assessed using numerical relativity simulations and extracting QNMs manually, {\it e.g.},  at late times by fitting a superposition of damped modes. 
However, if one wants to test general relativity, one should not rely on results of numerical relativity and look for QNMs as agnostic as possible! 
As outlined in section~\ref{sec3_linear_vs_nonlinear}, linear perturbation theory (which is necessary to define the QNM spectrum) may not be accurate enough to describe the ringdown at early times, independent of the presence of non-QNM contributions from the early response of the perturbed system. 
Possible general relativity violations that one could measure with high SNR events may thus be due to the limitations of perturbation theory itself. 
One possible effect is non-linear mode coupling, which can give rise to additional mode frequencies and damping times, which are however related to the first order QNMs. 
While these issues are still open problems in an idealized theoretical setup, real gravitational wave data is contaminated by complicated noise and events observed with current detectors only provide moderate SNR. 
Another complication is that the measured gravitational wave strain is a superposition of all possible modes, and thus it is more complicated label QNMs with their $(\ell, m, n)$. 

Claims of possible detections beyond the $\ell=m=2$ fundamental QNM are currently under discussion. 
In particular GW150914 provides a good start for spectroscopic tests due to the measured SNR of $\sim 25$ and suitable final black hole mass. 
For this event, the $n=1$ overtone of the $\ell=m=2$ QNM has been reported in references~\cite{Isi:2019aib}. 
However, this result has been questioned in reference~\cite{Cotesta:2022pci}, which points towards features in the noise. 
This possible explanation has been rejected by the authors of \cite{Isi:2019aib} in reference~\cite{Isi:2022mhy}. 
Another analysis has been carried out in reference~\cite{Finch:2022ynt} and reports tentative evidence in favor of the presence of the $n=1$ overtone, but with less significance as reported in reference~\cite{Isi:2019aib}. 
A possible detection beyond the $\ell=m=2$ fundamental QNM involving the $\ell=m=3$ fundamental QNM has been reported for GW190412 and GW190814 in reference~\cite{Capano:2020dix}. 
Another detection of the $\ell=m=3$ fundamental QNM has been reported for GW190521 in reference~\cite{Capano:2021etf}. 
A refined analysis by the same authors can be found in reference~\cite{Capano:2022zqm}. 

With the upcoming observing 4th run, one expects several hundreds of new sources, many of them with higher SNR. 
The subsequent observing run towards the late 2020's will be even more promising. 
Thus the question of whether QNMs beyond the $\ell=m=2$ fundamental QNM can be robustly inferred from data is only a matter of time~\cite{Cabero:2019zyt}, although one should not underestimate the overall complexity of rigorous and robust black hole spectroscopy ({\it e.g.}, \cite{CalderonBustillo:2020rmh}).
It may well be that even future detectors will be limited to a few QNMs, even under excellent conditions.

\subsection{Time Domain versus Frequency Domain Analysis}\label{sec5_TDvsFD}

Analyses of gravitational waves that focus uniquely on the ringdown have the advantage to directly connect with the predictions of black holes perturbation theory that we discussed in the previous sections. However, they also carry with them few disadvantages, that can be summarized by the fact that in order to isolate the ringdown from the rest of the signal, one needs to truncate it in some region close to where non-linearity is expected to be subdominant, or to match it with some description of the inspiral-merger phase. There are two different ways to tackle this problem which are orthogonal to each other and which we briefly explain in the next paragraphs: time domain and frequency domain analysis.

The LVK analysis pipeline of GW signals is almost fully carried out in the frequency domain. The reason is that the Fourier covariance matrix is diagonal and easily tractable. However, when one applies a Fourier transform to some truncated data, spectral leakage would start appearing and spoiling the analysis. The proposed solutions can be of two kinds: either applying some non-trivial windowing that regularizes the likelihood~\cite{Capano:2021etf,Zackay:2019kkv} or by matching the signal prior to the starting point either with some flexible fitting wave functions~\cite{Finch:2022ynt} or with semi-analytical models~\cite{Brito:2018rfr,Ghosh:2021mrv,Maggio:2022hre}.

One can avoid problems linked with spectral leakage with a time domain approach focusing only on the portion of the signal corresponding to the ringdown phase. The lack of a diagonal covariance matrix makes it necessary to properly treat the noise, which is evaluated through the auto-correlation function of the signal~\cite{Carullo:2019flw,Isi:2021iql}. This approach has been applied to the loudest events and had the problems discussed in the previous section, about a proper definition of the starting time.

Finally, a hybrid method based on {\it mode cleaning} of the signal has been proposed in reference~\cite{Ma:2023cwe,Ma:2023vvr}. 
In this method, one starts with a time domain likelihood, to which a rational filter is applied in the frequency domain. 
This is supposed to remove a single mode, and then the residual is analysed back in the time domain. 
According to the first results of its application to GW150914, which claim consistency with a first overtone in the signal, it looks like a very promising tool for future ringdown analysis.

\subsection{Constraints on Deviations from General Relativity}\label{sec5_theory_bounds}

\begin{figure}
\centering
\includegraphics[width=\linewidth]{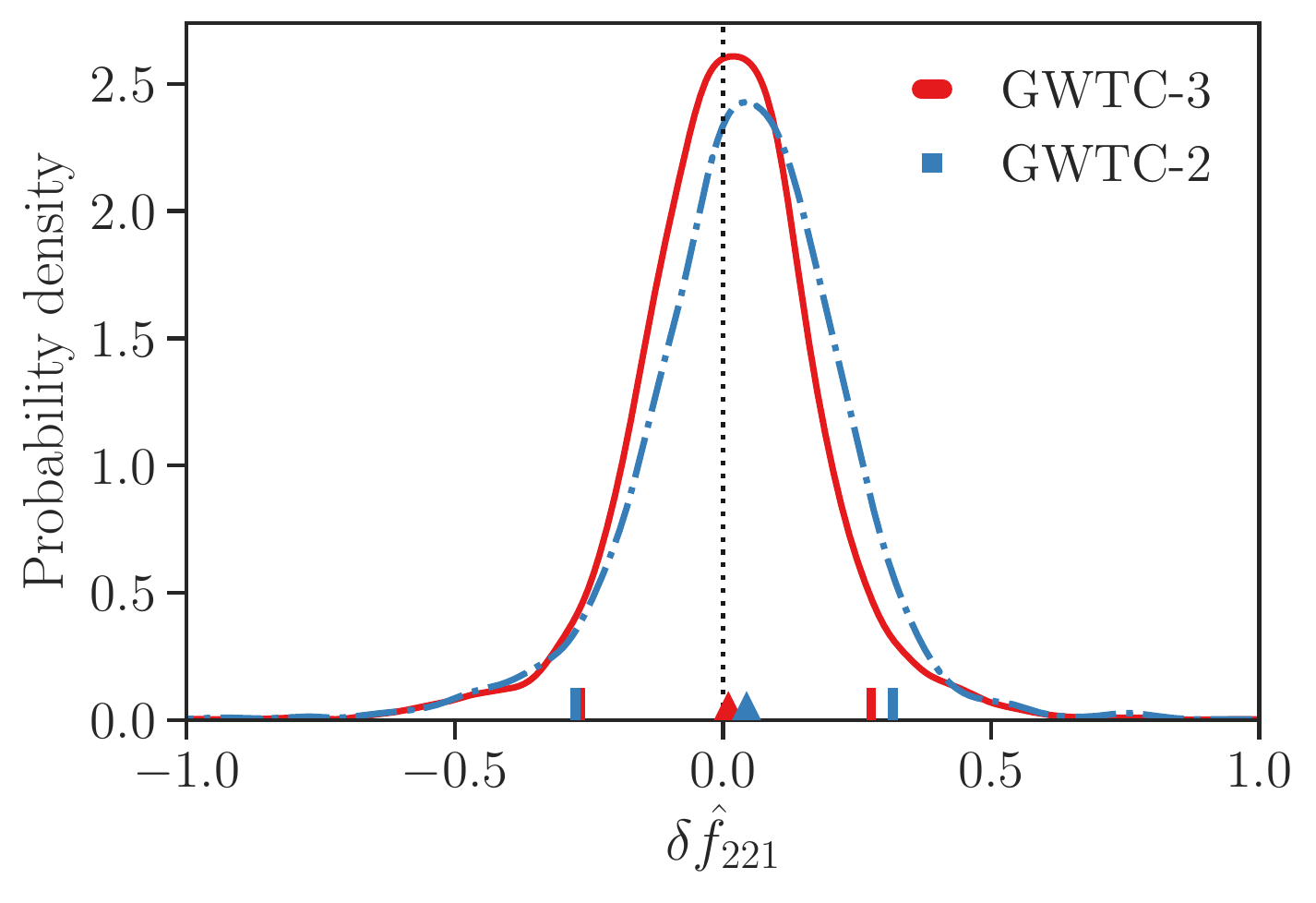}
\caption{Distribution of the probability density of the deviation of the real part of the $\ell=m=2$, $n=1$ mode accumulated on the various observation. Red solid line refers to the bound obtained considering all the events of the first three observing runs, while the blue dot-dashed line only includes those of the first two runs. Triangles mark the median of the distribution, and bars the 90\% confidence level. Credit: LVK Collaboration~\cite{LIGOScientific:2021sio} \href{https://creativecommons.org/licenses/by/4.0/}{CC BY 4.0}.}
\label{fig:ringdown_bounds}
\end{figure}

The debate on the detection of the first overtone and on higher modes does not hinder one from performing tests of general relativity with ringdown only data. 
Spectroscopy analysis was included in all LVK tests of general relativity papers~\cite{LIGOScientific:2016lio,LIGOScientific:2020tif,LIGOScientific:2021sio}. 
In the latest version, it included both time-domain and frequency-domain analysis. 
The study of the collaboration analysed the ringdown phase of 21 events among of all the binaries, with the possibility to select only those with sufficiently high SNR ratio. 
They tested blind deviations and found no evidence of beyond general relativity effects, as one can see from Fig.~\ref{fig:ringdown_bounds}.

After the release of data from the collaboration, further tests of general relativity came out. We discussed in section~\ref{sec4_agnostic_pheno} the existence of parametrized formalism that aim to describe more accurately how potential modifications in the gravity theory can affect the oscillation of the black hole. For example, Carullo analysed 3rd observing run data with {\it ParSpec}, showing that it can enhance the bounds that can be put on deviations, focusing on the class of theories which have a dimensional coupling (like sGB and dCS)~\cite{Carullo:2021dui}. An analysis which lead to similar conclusions was performed with the IMR frequency domain model by Silva {\it et al.}~\cite{Silva:2022srr}.

Attempts of theory-specific tests were also done. As we discussed in section~\ref{sec4}, the only non-Kerr rotating solution for which we have spectrum of QNMs is for Kerr-Newman black holes. Current data are not sufficient to place a sensible bound on an electromagnetic charge of black holes, unless the mass and spin are assumed to be known from inspiral and merger~\cite{Carullo:2021oxn}.

Another effect that we discussed in section~\ref{sec4} is the presence of extra polarizations due to the new couplings. The effect of possible non-tensor modes to the signal was first tested against current data in~\cite{Evstafyeva:2022rve} for the case of the scalar mode in sGB. Present data are too noisy to identify such mode, but projections for future detectors open the way to this possibility.

The general trend that one can infer from this handful of analyses for theory-specific and theory-agnostic approaches, is that ringdown is a promising tool for testing general relativity, but low-quality of current SNR ration makes it desirable to wait for ground based detector to reach higher sensitivity for beginning to put strong constraints or to show evidence of deviations. For example, in~\cite{Pacilio:2023mvk}, they injected deviations based on slow-spin extrapolation of several modified theories of gravity and showed with a Fisher analysis that small modifications are within the reach of future ground-based interferometers. Possible ways to improve this behaviour in the future are either by implementing a parametrized Bayesian framework, as suggested in~\cite{Meidam:2014jpa}, or by performing coherent mode stacking~\cite{Yang:2017zxs}.

\subsection{Status of Echo Searches}\label{sec5_echo}

Tentative evidence of echoes in gravitational wave observations has been claimed by several authors \cite{Abedi:2016hgu,Conklin:2017lwb,Abedi:2017isz,Abedi:2018npz,Abedi:2018pst,Holdom:2019bdv,Abedi:2020sgg}, but such findings were disputed by others \cite{Westerweck:2017hus,Lo:2018sep,Tsang:2019zra,Wang:2020ayy}. 
Also studies of the LVK Collaboration did not find evidence for echoes, but rather put constraints \cite{LIGOScientific:2020tif,LIGOScientific:2021sio}. 
In general it is very challenging to search for non-general relativity effects in existing observations for several reasons. 
A key limitation is the current sensitivity which requires a very careful treatment of the noise and implies that only the loudest part of the signal around the merger can be clearly identified. 
Thus a common problem is to quantify that any claimed evidence of echoes is not due to fitting features of the noise, which may not have any physical origin related to the source. 
Tightly related is the absence of a complete dynamical theory of ultra compact exotic compact objects which could be used to predict realistic waveforms to perform matched filtering in a similar way as for general relativity. 
Proposed templates and methods to construct them are rather phenomenological and allow to look for key features, {\it e.g.},  via forecasts with the Fisher matrix approach~\cite{Maselli:2017tfq,Testa:2018bzd}, but one cannot expect them to have similar accuracy as general relativity templates, which may make them less predictive. 
One exception are boson stars, for which mergers can be computed using numerical relativity techniques~\cite{Palenzuela:2017kcg,Bezares:2022obu}, but it is challenging to produce ultra compact remnants with echoes. 
For extreme mass ratio systems, which can be probed with future space born detectors (see section~\ref{sec6_future_detectors}), it is possible to generate basic waveforms with modified horizon properties using test particles as sources, see {\it e.g.}, reference~\cite{Maggio:2021uge}. 

\begin{svgraybox}
In this section we discussed:
\begin{itemize}
\item the LVK Collaboration has detected almost 100 compact binary mergers by the end of the third observing run;
\item robust detections of the $\ell=m=2$ fundamental QNM have been possible for several events, but claims for detection of the $\ell=m=2$ and $n=1$ overtone are debated, also for the $\ell=m=3$;
\item constraints on modified theories of gravity;
\item while some studies claim tentative evidence for gravitational wave echoes in existing data, other works including LVK studies do not.
\end{itemize}
\end{svgraybox}

\section{Outlook}\label{sec7}

The final section gives an outlook to the promising future of gravitational wave detectors in section~\ref{sec6_future_detectors}, discusses environmental effects in section~\ref{sec6_environmental} and outlines theoretical problems for QNMs of rotating black holes beyond general relativity in section~\ref{sec6_future_theory}.

\subsection{Future Gravitational Wave Detectors}\label{sec6_future_detectors}

The future prospects of observing the dark universe are bright! 
While the current gravitational wave detectors LVK will enhance their sensitivities by further upgrades and improvements in the upcoming years, the next generation of detectors like the Einstein Telescope~\cite{Maggiore:2019uih} and the Cosmic Explorer~\cite{Evans:2021gyd} are expected in the 2030's. 
These ground based detectors will operate in a similar frequency range as the current ones. 
Increased sensitivity will not only allow to observe similar events with higher SNR and thus allow for enhanced source studies and high precision tests of general relativity, but also increase the number of events by exploring a larger volume. 
This will be particular useful for population studies addressing questions such as the formation history and evolution of black holes. 

Complementary to these activities are the already ongoing measurements of pulsar timing arrays (PTAs), which work in a completely different way and frequency range. 
PTAs observe and combine several years of pulsar observations to detect very low frequency gravitational waves. 
Pulsars are neutron stars whose rotational period can be measured with high accuracy using radio telescopes. 
By measuring the delays between tiny deviations in the arrival times of the pulses, one can in principle measure gravitational waves. 
The frequency band that can be probed in this way is much lower than the one of other detectors and ranges roughly between $10^{-9}\,\mathrm{Hz}$ to $10^{-7}\,\mathrm{Hz}$. 
So far there have been no direct measurements of individual sources, {\it e.g.}, super-massive black hole inspirals, but tentative evidence for a stochastic background has been reported by the North American Nanohertz Observatory for Gravitational Waves (NANOGrav) Collaboration recently~\cite{NANOGrav:2020bcs}. 
The origin of this background is unclear at this moment, but it has since then been confirmed by the European Pulsar Timing Array (EPTA) Collaboration~\cite{Chen:2021rqp} and the International Pulsar Timing Array (IPTA) Collaboration~\cite{Antoniadis:2022pcn}. 
The so-called Hellings–Downs correlations~\cite{Hellings:1983fr}, which would indicate that the origin of this background is from gravitational waves, have not been found, at least so far. 
One should expect that ongoing efforts in the collection of pulsar data will track down the origin of the background in the future. 

In the mid 2030's there will be the ESA and NASA operated space-born gravitational wave detector LISA \cite{LISA:2017pwj}. 
After initial thoughts to work on such an experiment in the 1990's, huge technological advances and developments have been achieved. 
In recent years, the LISA pathfinder mission has been carried out very successfully \cite{Armano:2016bkm,Armano:2018kix}, which was a crucial step to achieve the ESA status for mission adoption. 
This implies that the mission will receive funding and planning to become reality in the mid 2030's. 
Plans for similar detectors have been proposed by China as TianQin~\cite{TianQin:2015yph,TianQin:2020hid} and Taiji~\cite{Hu:2017mde,Ruan:2018tsw}. 
The observable frequency range of these detectors will be between PTAs and ground based detectors, thus aiming for intermediate mass and supermassive black holes. 
In the context of black hole spectroscopy, it will provide very high SNR signals with unprecedented accuracy~\cite{Berti:2005ys}. 
From a fundamental physics point of view, the list of open problems related to the science case is still immense and calls for much progress in the years ahead~\cite{Barausse:2020rsu}. 
Among the new expected sources compared to current detectors are extreme mass ratio inspirals (EMRIs). 
Here a supermassive black hole is orbited by a much lighter companion, {\it e.g.},  stellar mass black holes or neutron stars. 
Besides uncertain predictions for the expected rate of such events, the very long signal duration makes current modeling techniques insufficient for accurate waveform modeling. 
However, recent progress in the field of gravitational self-force seems very promising, {\it e.g.},  second-order results for non-spinning, quasi-circular inspirals exist~\cite{Wardell:2021fyy}, although the computational challenges for more general cases are very involved. 
If EMRI signals can be modeled accurately, they may also provide evidence with respect to tests of general relativity for multiple reasons. 
EMRIs allow, in principle, to map the space-time of the supermassive black hole, which under common assumptions, should be described by the Kerr metric. 
One possible effect that could appear if the background metric is not the Kerr metric, is the presence of chaotic dynamics and frequency jumps in the waveform, as shown in reference~\cite{Destounis:2021mqv}. 
Possible beyond general relativity effects that have been studied in the context of EMRIs also arise in theories with additional scalar fields, whose presence would add an additional channel of loss of energy, and would leave a detectable imprint in the signal~\cite{Maselli:2020zgv,Maselli:2021men,Barsanti:2022ana}. 
However, any claim of beyond general relativity signals will require extraordinary evidence, as there is also a long list of possible environmental effects, which could alter the signal and thus be falsely attributed. 

Another direction of open problems is about the challenges of modeling many sources simultaneously, which is also known as the so-called ``global fit'' problem~\cite{Cornish:2005qw,Vallisneri:2008ye,MockLISADataChallengeTaskForce:2009wir}. 
In contrast to current detectors, where it is very unlikely that signals contain more than one merger event at a time, the time-scales for the sources in the LISA band are very different. 
Many of these sources will emit gravitational waves from an early stage of the inspiral and will not merge within the entire lifespan of the mission. 
While this allows unprecedented mapping of compact objects, {\it e.g.},  the white dwarf binary population in our galaxy, it also introduces new challenges to precision tests of general relativity due to the outstanding complexity in understanding all systematic effects. 

Finally, one exciting future possibility is to complement QNMs constraints with insights gained with next generation EHT experiments. 
At first the two sides seem disconnected, and it is not expected that measurements of the same black hole can be made in both approaches, but images of black holes and properties of QNMs are related with each other, {\it e.g.}, references~\cite{Jusufi:2020dhz,Yang:2021zqy} discuss the relation for the Kerr black hole. 
The characteristic ring feature known from existing EHT images is directly related to the impact parameter of the photon ring, which in the eikonal limit, is also related to QNMs. 
Thus, constraints on QNMs can in principle provide complementary information on the black hole metric at the photon ring\footnote{For instance, see Fig.~2 in reference~\cite{Volkel:2020xlc}.}. 
First bounds on beyond general relativity metrics, theory specific and agnostic ones, have already been discussed for M87* and/or Sgr A*, {\it e.g.}, in Refs~\cite{Cunha:2019ikd,EventHorizonTelescope:2020qrl,Volkel:2020xlc,EventHorizonTelescope:2021dqv,EventHorizonTelescope:2022xqj}. 
While the full image is much more prone to degeneracies between modified metrics and uncertain accretion physics~\cite{Gralla:2020pra,Lara:2021zth}, next generation EHT observations might be able to further explore the shape of the main ring~\cite{Gralla:2020srx} and a subring structure that could be used as more robust features to explore the underlying black hole metric~\cite{Ayzenberg:2022twz}. 
However, it is also known that non-Kerr metrics can in principle still reproduce similar results, which can still leave some room for degeneracy, see. {\it e.g.}, references~\cite{Glampedakis:2021oie,Herdeiro:2021lwl,Glampedakis:2023eek}.

\subsection{Environmental Effects on Quasi-Normal Modes}\label{sec6_environmental}

All our previous discussions and also those in the subsequent sections focus on QNMs of vacuum black holes. 
In contrast to electromagnetic radiation, gravitational waves propagate almost without any absorption or scattering through astrophysical environments. 
Moreover, for binary black hole mergers observable with current detectors, one would not expect significant amounts of matter in their close vicinity, especially not in the final inspiral phase. 
This expectation seems in agreement with the lack of any electromagnetic counter part for binary black hole mergers so far. 

However, there are several scenarios in which non-vacuum environments could have measurable effects which would challenge the correct analysis and interpretation of gravitational wave signals. 
Most of these scenarios may be relevant for some of the sources observable future detectors like LISA and are rather secular effects that grow over time. 
Some binary black hole systems, which would be in a much earlier stage of their inspiral long before merger and thus in the LISA frequency band, could evolve in a gaseous environment. 
This introduces additional drag and friction mechanisms which could shrink the binary orbit additionally to the emission of gravitational waves. 
Such effect could lead to biased source parameters, if it is significant enough and if one assumes a vacuum environment. 
Other effects could be due to a binary black hole system being perturbed by a third compact object, {\it e.g.},  in globular clusters or in the vicinity of super massive black holes. 

The situation for QNMs is somehow different.
Because they are typically only excited through the merger and characterize the perturbations in the vicinity of the final black hole, one may think they cannot be impacted. 
However, there can be quite drastic effects on the QNM spectrum if small, but non-zero matter distributions are present. Somewhat surprisingly at first glance, these effects are typically more drastic the further away the matter is. 
This is closely related to the significance of QNMs discussed in section~\ref{sec3_relevance}. 
In the following we highlight some of the simple cases and refer the reader to reference~\cite{Barausse:2014tra} for a detailed review on environmental effects on gravitational-wave astrophysics for more details. 

As simplest approach to the problem one can consider scalar perturbations on a background space-time that approximates a black hole surrounded by a spherical shell of matter, as done in reference~\cite{Barausse:2014tra}. 
Although this is not adequate to quantify realistic effects, it allows to study possible main effects without going through the significantly more involved exercise of solving the linearized Einstein field equations with  source term. 
As it turns out, in this case the presence of the shell only modifies the effective potential term by terms including the radial mass distribution of the shell and its first derivative. 
This problem can then be studied using suitable techniques presented in section~\ref{sec3_methods_qnms}, {\it e.g.},  the shooting method to obtain the QNM spectrum and the time evolution to study the evolution of perturbations. 

The overall finding, which has also recently been revisited in more detail with the so-called pseudo-spectrum method~\cite{Jaramillo:2020tuu,Jaramillo:2021tmt,Cheung:2021bol,Boyanov:2022ark} and is in agreement with the pioneering works of Nollert~\cite{Nollert:1996rf}, and Nollert and Price~\cite{Nollert:1998ys}, is that small perturbations in the potential can introduce significant changes in the QNM spectrum, especially for the rather unique structure of black hole overtones. 
See also reference~\cite{Konoplya:2022pbc} for a recent study on how overtones of parametrized black hole metrics deviate stronger for overtones, which is in agreement with complementary expectations from the parametrized QNM framework \cite{Volkel:2022aca}. 
For a detailed and quantitative discussion we refer to the previously mentioned references, but some aspects can also be qualitatively understood. 
For example, if the modification of the potential corresponds to a localized bump far away from the angular momentum barrier, it is possible that low frequency waves get trapped between the two bumps and thus change the structure of the eigenvalue spectrum. 
The further the small bump is away, the easier it is to trap low frequency waves. 
Does this imply the QNM spectrum is practically irrelevant for the non-vacuum case? 

To address this question one has to consider the actual evolution of perturbations, as this is more appropriate to what one could observe with gravitational wave detectors far away from the source. 
Without going into too much details about specific configurations that might further complicate the overall picture (again we refer to the previous studies), one finds that the early response is dominated by the properties of the non-perturbed potential and thus mimicks the original ringdown and thus mainly the fundamental QNM. 
However, as outlined in section~\ref{sec4}, time-domain calculations for black holes only probe the fundamental mode and very few overtones. 
At later times, it is well possible to observe some of the true QNMs of the perturbed potential, {\it e.g.},  when they correspond to quasi-trapped low frequency modes. 

The study of environmental effects on QNMs with perturbation theory is mostly limited to spherical symmetry and static configurations, which may not reflect realistic environments. 
Because future detectors will provide high SNR observations, even small effects could become important and more quantitative studies are certainly needed. 

As final remark we want to briefly mention gravitational lensing~\cite{Bartelmann:2010fz}. 
It is well known that gravitational fields of galaxy clusters or other massive sources do not only impact the propagation of light, but of course also the one of gravitational waves. 
Although there is so far no evidence for lensing effects on existing gravitational wave observations~\cite{LIGOScientific:2021izm,LIGOScientific:2023bwz}, it is well possible that at least some future signals will be impacted. 
If the effects are strong enough and are not modelled, they could introduce systematic effects in the standard data analysis and might indicate violations of the Kerr hypothesis due to a modified ringdown spectrum.

\subsection{QNMs of Rotating Black Holes in Alternative Theories}\label{sec6_future_theory}

The current status of the computation of QNMs for rotating black holes beyond general relativity has been discussed in detail in section~\ref{sec5_overview_theories}. We reported how this is a currently open line of research, were some advancement has been done, yet we are still far to have a complete understanding of the problem. Anyway, it is very likely that in the next forthcoming years we will be able to fill this gap, initially with slow-rotation or small coupling expansions, and eventually with full numerical solution of the equations.
We want to emphasize that knowledge of spectra of rotating black holes in specific theories is necessary if we want to be able to set constraints on such theories with ringdown observations, as well as performing a Bayesian comparison with general relativity. 

A big caveat to this approach is that it turned out to be valid for some theories and/or solutions, but some of them have properties for which the study of perturbation theory is extremely complicated. One emblematic case is represented by Ho\v rava and Einstein-{\AAEE}ther theory, which broadly fall within the class of Lorentz-violating gravity. With the exception of a specific subclass of Ho\v rava gravity, for which perturbations are identical to those of general relativity~\cite{Franchini:2021bpt}, there is not any universally common understanding of the problem even for the spherically symmetric case. In this class of theories, one can find new degrees of freedom on top of the gravitational one, with the special property that, since the theory violates Lorentz invariance, they can propagate faster than light. The main issue is that no serious proposal has been made about the boundary conditions that one would need to set for the various degrees of freedom at the innermost boundary of the spacetime, which is given by the universal horizon, a causal one-directional-in-time membrane acting for any degree of freedom~\cite{Barausse:2011pu,Blas:2011ni}.

\begin{svgraybox}
In this section we discussed:
\begin{itemize}
\item future gravitational wave detectors like the Einstein Telescope and Cosmic Explorer (current frequency range), LISA and TianQin as space-born low-frequency detectors, and ongoing activities of pulsar timing arrays;
\item scenarios in which environmental effects can influence the QNM spectrum, like shells of matter or gravitational lensing;
\item overview and necessary work for QNMs of rotating black holes beyond general relativity.
\end{itemize}
\end{svgraybox}

\begin{acknowledgement}
The authors want to thank Enrico Barausse, Kostas Glampedakis, Roman Konoplya and Laura Sberna for providing useful discussions and comments that enriched this chapter. 
S.~H.~V\"olkel acknowledges funding from the Deutsche Forschungsgemeinschaft (DFG) - project number: 386119226. 
\end{acknowledgement}

\bibliographystyle{myspringer}
\bibliography{literature}

\end{document}